\documentclass[lettersize,journal]{IEEEtran}
\usepackage{amsmath,amsfonts}
\usepackage{algorithm}
\usepackage{array}
\usepackage{textcomp}
\usepackage{stfloats}
\usepackage{url}
\usepackage{verbatim}
\usepackage{graphicx}
\usepackage{cite}

\usepackage{tabularx}
\usepackage{subfigure}
\usepackage{bm}
\usepackage{algpseudocode}
\usepackage{algorithmicx}
\usepackage{color}
\usepackage{amsmath}
\usepackage{mathrsfs}
\usepackage{amsfonts}
\usepackage{hyperref}
\usepackage{comment}
\usepackage{upgreek}
\usepackage{ulem}

\begin{document}

\title{Improving Adaptive Real-Time Video Communication Via Cross-layer Optimization}

\author{Yueheng~Li,
        Hao~Chen,~\IEEEmembership{Member,~IEEE,}
        Bowei~Xu,
        Zicheng~Zhang,
        and~Zhan~Ma,~\IEEEmembership{Senior~Member,~IEEE}
\thanks{All authors are
with Nanjing University, Nanjing 210023,
China. E-mails: yueheng.li@smail.nju.edu.cn, chenhao1210@nju.edu.cn, mariogotze@163.com, zichengzhang@smail.nju.edu.cn, mazhan@nju.edu.cn.}
}

\markboth{IEEE TRANSACTIONS ON MULTIMEDIA,~Vol.~XX, No.~X, August~2022}%
{Li \MakeLowercase{\textit{et al.}}: Improving Adaptive Real-Time Video Communication Via Cross-layer Optimization}

\IEEEpubid{0000--0000/00\$00.00~\copyright~2022 IEEE}

\maketitle

\begin{abstract}
Effective Adaptive Bitrate (ABR) algorithm or policy is of paramount importance for Real-Time Video Communication (RTVC)  amid this pandemic to pursue uncompromised quality of experience (QoE). Existing ABR methods mainly separate the network bandwidth estimation and video encoder control, and fine-tune video bitrate towards estimated bandwidth, assuming the maximization of bandwidth utilization yields the optimal QoE. However, the QoE of an RTVC system is jointly determined by the quality of the compressed video, fluency of video playback, and interaction delay. Solely maximizing the bandwidth utilization without comprehensively considering compound impacts incurred by both network and video application layers, does not assure a satisfactory QoE. And the decoupling of the network and video layer further exacerbates the user experience due to network-codec incoordination. This work, therefore, proposes the Palette, a reinforcement learning-based ABR scheme that unifies the processing of network and video application layers to directly maximize the QoE formulated as the weighted function of video quality, stalling rate, and delay. To this aim, a cross-layer optimization is proposed to derive the fine-grained compression factor of the upcoming frame(s) using cross-layer observations like network conditions, video encoding parameters, and video content complexity. As a result, Palette manages to resolve the network-codec incoordination and to best catch up with the network fluctuation. Compared with state-of-the-art schemes in real-world tests, Palette not only reduces 3.1\%-46.3\% of the stalling rate, 20.2\%-50.8\% of the delay but also improves 0.2\%-7.2\% of the video quality with comparable bandwidth consumption, under a variety of application scenarios.
\end{abstract}

\begin{IEEEkeywords}
Adaptive bitrate, cross-layer optimization, network condition, video encoding parameter, and video content complexity.
\end{IEEEkeywords}

\section{Introduction} \label{sec:intro}



\IEEEPARstart{R}{ecent} years have witnessed the exponential growth of real-time video communication (RTVC) system-based applications, e.g., online education, remote sharing, video conferencing, etc.  Particularly, since the outbreak of COVID-19, the use of cloud-based video conferencing has increased about 5$\times$ due to travel and social distance restrictions~\cite{market_report2020}. These RTVC systems are now widely used in our daily life for entertainment and for work, leading to significant Internet video traffic consumption~\cite{video_traffic2022}  (e.g., $>17\%$) and multi-billion revenue return worldwide~\cite{video_market2022}. 

To ensure smooth service provisioning with uncompromised quality of service (QoE), numerous adaptive bitrate (ABR) methods have emerged to deal with unexpected fluctuation of underlying heterogeneous access networks (e.g., WiFi, 4G, 5G, etc) in RTVC applications. More importantly, built upon the powerful representation capacity of deep learning techniques, various learning-based ABR decision approaches have been developed with the focus on RTVC service enabling like cloud gaming~\cite{chen2019tgaming,8685768}, mobile video telephony~\cite{zhou2019learning}, e-commerce live shows and virtual interactive shopping~\cite{Loki}, quantitatively exemplifying noticeable performance improvement on average, e.g., higher network bandwidth utilization, less delay (in ms), etc., against default rules-based Google Congestion Control (GCC)~\cite{10.1109/TNET.2017.2703615} used in numerous products.

\subsection{Status Quo of Existing ABR Solutions}

Optimizing ABR policy for Internet-scale RTVC systems is complex and challenging. Thus, researchers and practitioners often resort to the principle of "Divide and Conquer", where a big problem is first broken into smaller and solvable modules for easy implementation, fast validation, and robust component reuse~\cite{952802}. As a result, existing rules-based and learning-based ABR solutions mostly decouple the estimation of available network bandwidth (e.g., bandwidth target) and video encoder rate control (e.g., compressed video bitrate), and tune video bitrate towards bandwidth target by simply assuming that the video quality received at client side is highly correlated with the video bitrate.

However, the QoE at end users in RTVC systems is comprehensively 
impacted by a variety of factors from both network and application layers, rather than just a simple video bitrate. For instance, network congestion may lead to video stalling that greatly deteriorates the perceptual sensation~\cite{frame_stalling_quality}. In the meantime, as analytically deduced in~\cite{hu2012optimization}, the quality of compressed video at the sender side is jointly related to the bitrate and its content complexity. Having the same bitrate for motion-intensive cloud gaming and stationary talking-head conferencing would give many different sensations of video quality to the end user~\cite{ma2011modeling}. Though few ABR explorations in~\cite{Loki,chen2019tgaming,OnRL} have looked into the stalling impact on the QoE, compound impacts from cross-layer factors are still largely overlooked in rate-oriented methods, yielding inferior QoE with large stalling rate 
and high delay (see Fig.~\ref{fig:contents}). 

On the other hand, after obtaining the available bandwidth target, most existing RTVC systems solely rely on the video encoder to generate compressed video with a bitrate as close to the \IEEEpubidadjcol target as possible. However, network-codec incoordination (i.e., network bandwidth estimation can run at millisecond-scale while the video encoder requires at least one second and even several seconds to stabilize and match the target bitrate~\cite{zhou2019learning}), often leads to notable ``adaptation lag'' that incurs bitrate overshooting or undershooting against the bandwidth target from time to time, resulting in frequent occurrences of video stalling and high delay (see Fig.~\ref{fig:algs}).

\subsection{Our Method}

This work therefore, proposes the Palette\footnote{A palette is used to mix a variety of colors  to generate a better one on purpose which is more or less the same as the proposed approach that jointly combines the states of cross-layer factors to best determine fine-grained compression factor for optimal QoE.}, to determine appropriate fine-grained compression factor, e.g., constant rate factor (CRF),
for upcoming frames, by jointly considering the network conditions (e.g., packet loss rate, delay measured by the round trip time (RTT), and stalling rate), video encoding parameters (e.g., frame type, CRF), and video content complexity measured by the spatial perceptual information (SI) and temporal perceptual information index (TI)~\cite{ITU-R_BT.1788} in the past, through a reinforcement learning (RL) based approach. 

Further, a multiscale strategy is directed to facilitate fine-grained frame-level rate control in which the aforementioned average CRF for a group of temporally-successive frames is used to determine per-frame quantization parameter (QP) by underlying video encoder accordingly. By applying this CRF rate control mode, the encoder optimizes the perceptual quality of compressed video with pleasant spatial and temporal coherency~\cite{ou2010perceptual} towards the optimal QoE.

As seen, Palette unifies the network bandwidth estimation and video encoding adaptation, with which its frame-level video bitrate adjustment can best alleviate the aforementioned ``adaptation lag'' issue by promptly responding to the network bandwidth fluctuation from time to time (see Fig.~\ref{fig:algs}). 




To evaluate the performance of Palette, we develop the {\it Echo}\footnote{We develop {\it Echo} based on an open-source WebRTC framework which is available at \url{https://github.com/yuanrongxi/razor}.} - a WebRTC~\cite{webrtc,webrtc_standard} based real-time video conferencing testbed as a typical RTVC application. In addition to Palette, three other state-of-the-art ABR approaches are deployed on the same testbed for a fair comparison, including rules-based GCC~\cite{10.1109/TNET.2017.2703615}, 
reinforcement learning (RL) based Adaptive Real-time Streaming (ARS)~\cite{chen2019tgaming}, and imitation learning (IL) based Concerto~\cite{zhou2019learning}. 
Both lab-managed trace simulations and real-world field trials have revealed the superior performance of Palette. In trace-driven tests, Palette outperforms the state-of-the-art algorithms with a reduction of 26.1\%-65.3\%, and 39.5\%-55.1\% for respective stalling rate and delay on average; and saves 12.3\%-15.2\% bandwidth at the same video quality measured by prevalent VMAF index (Video Multi-Method Assessment Fusion)~\cite{VMAF}. As for the real-world tests held in a variety of application scenarios, Palette not only reduces 3.1\%-46.3\% of the stalling rate, 20.2\%-50.8\% of the delay but also improves 0.2\%-7.2\% of the video quality with the similar bandwidth consumption.

%

\subsection{Contributions} The proposed Palette makes the following contributions:
\begin{itemize}
    
    \item This work unifies the processing of network and video application layers to drive frame-level encoder adaption, which effectively resolves the network-codec incoordination to better catch up with the network fluctuations;
    \item This work determines the averaged CRF for a few successive frames by jointly considering the compound impacts of QoE-oriented cross-layer factors, including the network conditions, video encoding parameters, and video content complexity towards the best QoE but not the maximization of network utilization;
    
    \item Both lab-managed simulations and real-world field tests in the wild report the superior performance of Palette to the state-of-the-art.
\end{itemize}

\section{Related Work}

Recent years have seen a rapid increase of ABR approaches aiming to optimize the QoE in various networked video scenarios like Video on Demand (VoD) and RTVC. Representative examples using either rules-based or learning-based solutions are briefed for both applications as follows. Whereas, the development of the ABR approach for RTVC services is more challenging since a multi-second playback buffer used in VoD applications for mitigating network fluctuations is mostly prohibited and the actual compressed video bitrate is often unavailable to the ABR engine until completing the video encoding session.


{\bf Rules-based approaches} typically predict future network bandwidth according to a sequence of observed states in the past from network layers like playback buffer size (if applicable), delay, packet loss, and throughput. For VoD services, the most representative example is the MPC (Model Predictive Control)~\cite{MPC}, which determines the bitrate to stream the next chunk based on throughput estimation and playback buffer size in the past. For RTVC services, the GCC is a prevalent solution in use~\cite{10.1109/TNET.2017.2703615}, which predicts the future bandwidth using both end-to-end delay variations and packet loss rates observed from preceding slots. Now, it is used in WebRTC standard~\cite{webrtc,webrtc_standard} as the default congestion control algorithm that has been overwhelmingly deployed in billions of devices.




{\bf Learning-based approaches} have shown their superior performances recently. Well-known learning-based models for applications with sufficient playback buffer support like VoD or HTTP-based live streaming are Pensieve~\cite{Pensieve}, RLVA~\cite{9226435}, QARC~\cite{QARC}, Oboe~\cite{Oboe}, and Comyco~\cite{Comyco}. Most of them simply tune the video bitrate to maximize bandwidth utilization, while the QARC suggests mapping the video bitrate to the corresponding VMAF score for quality optimization. 
Maximizing network utilization is also extended for latency-sensitive RTVC services as in Concerto~\cite{zhou2019learning}, which uses imitation learning to aggressively match the available network bandwidth. In the meantime, RL-based ARS~\cite{chen2019tgaming} considers the encoding bitrate together with the fluency of video playback for optimization.


Recently, the combination of rules-based and learning-based approaches for ABR decisions attracts intensive attention, for which it wishes to leverage the advantages from both aspects. For instance, Stick~\cite{Stick} improves buffer-based ABR algorithms for VoD streaming by adjusting its buffer bounds, which are determined by an additional RL model. As for RTVC applications, OnRL~\cite{OnRL} applies robust hybrid learning in which it would switch to the GCC when RL based algorithm runs abnormally. Later, Loki~\cite{Loki} proposes a deeper fusion scheme that allows the learning-based and rules-based approaches to work synchronously to co-determine a video bitrate for maximizing bandwidth utilization.\footnote{{Both OnRL and Loki are not open-source projects and the performance results presented in their paper are evaluated in a customized commercial system.}}

As extensively studied in~\cite{ma2011modeling,frame_stalling_quality,hu2012optimization}, the QoE of networked video is compound impacted by cross-layer factors like stalling rate, delay, bitrate, and content complexity. However, existing bitrate adaption approaches for RTVC services, including rules-based (e.g., GCC), learning-based (e.g., Concerto, ARS), and hybrid (e.g., OnRL, Loki), are still estimating the network bandwidth only and other QoE factors are largely overlooked. On the contrary, this work resorts to the QoE optimization by comprehensively considering various factors aforementioned for adaption decisions.


{\bf The codec-transport incoordination} was first reported in Salsify~\cite{211251}. To tackle it, Salsify proposed a per-frame bitrate adaptation scheme to quickly respond to network fluctuations. However, this requires the customization of video codec, making Salsify incompatible with general protocols. Though extensive studies had been also carried out to analyze such codec-transport incoordination in Concerto~\cite{zhou2019learning}, it did not come up with a radical solution since it still fully relied on the video encoder to perform rate control, and the adaptation lag is inevitably presented (see Fig.~\ref{fig:algs}). Such codec-transport incoordination is mainly due to the decoupled processing of network bandwidth estimation and video encoder rate control. This work, therefore, unifies the processing of network and video application layers to perform fine-grained frame-level compression factor determination to resolve the codec-transport incoordination problem fundamentally and to improve the overall QoE.


 
 
 
\section{Measurements and Observations}
\label{sec:observation}
In this section, we perform exhaustive measurements to examine existing ABR approaches used in RTVC services and to understand their limitations.


\subsection{Measurement Setup} \label{sec:measurement_setup}
Three prevalent ABR approaches, e.g., GCC~\cite{10.1109/TNET.2017.2703615}, ARS~\cite{chen2019tgaming} and Concerto~\cite{zhou2019learning} are deployed on {\it Echo} to study their limitations. We run {\it Echo} with these ABR solutions in a lab-managed environment for simplicity. To best emulate the real-world network behaviors, we generate typical trace segments by synthesizing real network traces (e.g., HSDPA~\cite{10.1145/2483977.2483991}, Oboe~\cite{Oboe}, FCC~\cite{fcc2016}, and MMGC2019~\cite{mmgc2019}). We then apply the widely-used Linux Traffic Control (TC) tool~\cite{linux_tc} to simulate the network following the synthetic trace segments. More details regarding the {\it Echo} and network traces are given in Sec.~\ref{ssec:eva_Methodology}.


In this measurement, {\it Echo} are installed on two personal computers (PCs) that are inter-connected using an Ethernet switch, where one is set as the video sender and the other is the receiver. The sender encodes the live scenes rendered from pre-cached video files and streams them to the receiver for consumption in real time. We use a variety of video content to simulate different application scenarios (e.g., conferencing, gaming, etc.).

\begin{figure}[t]
    \centering
    \includegraphics[width=\linewidth]{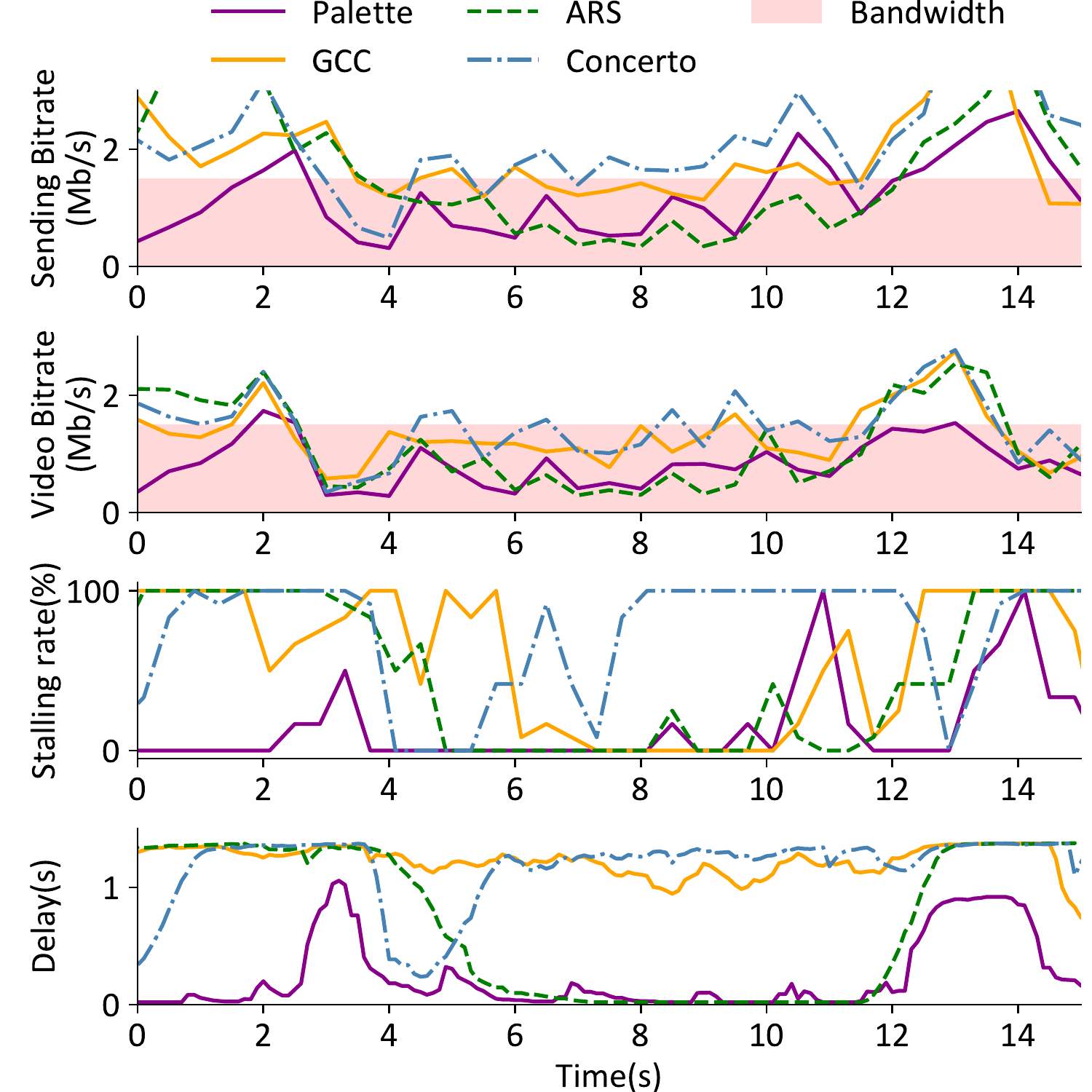}
    \caption{\color{black}{\it Maximizing Bandwidth Utilization Does NOT Assure Optimal QoE.} Obvious video bitrate overshooting induced video stalling and high delay are presented for GCC~\cite{10.1109/TNET.2017.2703615} and Concerto~\cite{zhou2019learning} that primarily optimize the bandwidth utilization. Better performance is obtained using ARS~\cite{chen2019tgaming} which includes the stalling rate during optimization. {Palette shows the lowest stalling rate (defined as the ratio of stalling time to total playback time) and smallest delay (measured by RTT)} by jointly considering cross-layer QoE factors. A gaming video with a dynamic scene is experimented with a fixed bandwidth to avoid network fluctuation induced intervention.}
    
    
    \label{fig:contents}
\end{figure}

\subsection{Observations} \label{ssec:observations}

\subsubsection{Maximizing Bandwidth Utilization Does NOT Assure Optimal QoE} Maximizing the QoE is a constant pursuit of an RTVC application for its success. Prevalent ABR approaches choose to tune the video bitrate towards the bandwidth target (a.k.a., the maximization of network bandwidth utilization) under the simple assumption that a higher video bitrate comes with better video quality and QoE. 

However, as jointly determined by the compressed video quality, the fluency of video playback, and interaction delay, the QoE in RTVC services depends on a variety of factors like video content complexity, stalling rate, etc. For example, encoding a video with dynamic complex content would present drastic changes in video bitrate, which easily causes the bitrate overshooting issue.
In the meantime, simply pursuing the higher bandwidth utilization further increases the risk of bitrate overshooting in RTVC scenarios that exhibit typical traffic bursts~\cite{zhou2019learning}. The occurrences of bitrate overshooting would generally incur video stalling and high delay events, severely deteriorating the QoE.

To better understand this problem, we run the GCC, ARS, and Concerto using a first-person shooting game video with high dynamic scenes under a fixed-bandwidth (e.g., 1.5Mbps) network trace. Having a constant bandwidth allows us to avoid potential impacts incurred by the network fluctuation, and to focus on the other QoE-oriented factors. The resulting performances are plotted in Fig.~\ref{fig:contents}. {\color{black} As seen, higher video bitrate is presented for GCC and Concerto, yielding obvious sending bitrate overshooting induced severe video stalling and high delay (e.g., from 6s to the end).} This is because GCC and Concerto solely tune the bitrate to maximize bandwidth utilization. Later, as the ARS~\cite{chen2019tgaming} considers the stalling rate for optimization besides the bitrate, it clearly shows the reduction of the stalling rate and delay from 5s to 10s. But the ARS also fails to avoid the bitrate overshooting at around 2s and from 12s to 14s where fast scene changes are presented in the test video.

As seen, only tuning the video bitrate to maximize the bandwidth utilization is insufficient to assure the uncompromised QoE. {\it A better way is to directly optimize the QoE with full consideration of compound impacts from network and video application layers.} 

\begin{figure}[t]
    \centering
  \includegraphics[width=\linewidth]{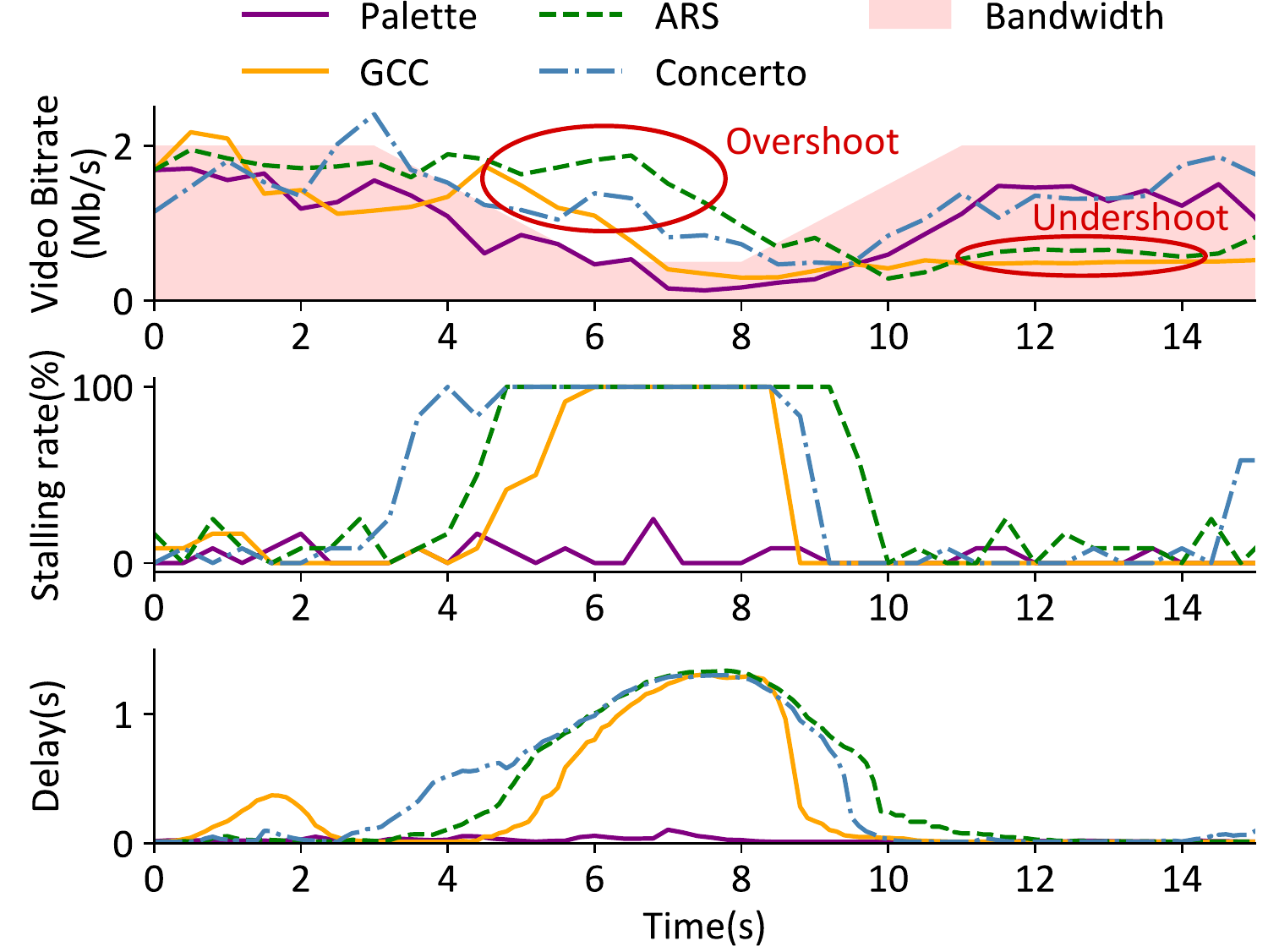}
    \caption{{\it Adaptation Lag.} Video bitrate overshooting is generally appeared in GCC~\cite{10.1109/TNET.2017.2703615}, ARS~\cite{chen2019tgaming}, and Concerto~\cite{zhou2019learning} because of the network-codec incoordination, which incurs the significant increase in stalling rate and delay. By contrast, the proposed Palette unifies the network and video application layers to adapt frame-level CRF, by which it can promptly respond to network fluctuations with much less stalling and smaller delay. A video with stationary content is used to just study the network fluctuation impact.} 
    \label{fig:algs}
\end{figure}

\subsubsection{Adaptation Lag}

In existing ABR approaches, network bandwidth estimation, and video encoder control modules are decoupled separately, implying that we solely rely on the video encoder to perform the rate control~\cite{10.1109/TNET.2017.2703615,chen2019tgaming,zhou2019learning} after obtaining the bandwidth target. However, network-codec incoordination~\cite{211251} leads to notable ``adaptation lag'' and severe QoE degradation.

To fully understand the impact of ``adaptation lag'', we first generate a synthetic network trace segment that best mimics typical network fluctuations in practice. As shown in the top part of Fig.~\ref{fig:algs}, the available bandwidth gradually drops from 2Mbps to 0.5Mbps from 3s to 6s and then increases at around 8s until the complete recovery at 11s.  

As revealed, the adaptations of video bitrate in respective GCC, ARS, and Concerto clearly lag behind the network fluctuation, resulting in bitrate overshooting when the network bandwidth drops and a sharp increase of stalling rate and delay at the similar time shown in the middle and bottom part of Fig.~\ref{fig:algs}. Such lag also appears when the network gradually recovers, leading to bitrate undershooting.


We notice that, though adaptation lag-induced bitrate over- and under-shooting always appear, the compressed video bitrate in Concerto~\cite{zhou2019learning} is relatively closer to the actual network bandwidth in comparison to that of GCC and ARS. Since we are using the same encoder on {\it Echo}, this result implies that the Concerto can offer a better estimation of available network bandwidth.

In summary, it suggests that {\it a fine-grained video bitrate control is highly desired to instantaneously respond to network dynamics without lag.}

\section{Palette Design}
\label{sec:design}

{\color{black} To address the two critical problems of existing ABR schemes aforementioned in Sec.~\ref{ssec:observations}, Palette is designed to optimize end-to-end QoE in both the application and transport layers, which is summed up as avoiding video stall and delay while maximizing picture quality. Palette achieves fine-grained adjustment of the actual video bitrate by directly deciding encoding parameters (e.g., CRF) at the frame level, avoiding the ``adaptation lag'' caused by the long convergence period of the inherent bitrate control of the codec (Sec.~\ref{ssec:system_overview}). Additionally, instead of maximizing bandwidth utilization, Palette directly optimizes the QoE by fully considering QoE-oriented cross-layer factors observed from the network and video application layers (Sec.~\ref{ssec:factors}). Palette utilizes deep reinforcement learning (DRL) to generate the end-to-end ABR policy which is represented by a neural network (Sec.~\ref{ssec:drl_model}). }

\subsection{System Architecture}
\label{ssec:system_overview}

\begin{figure}[t]
    \centering
    \includegraphics[width=0.9\linewidth]{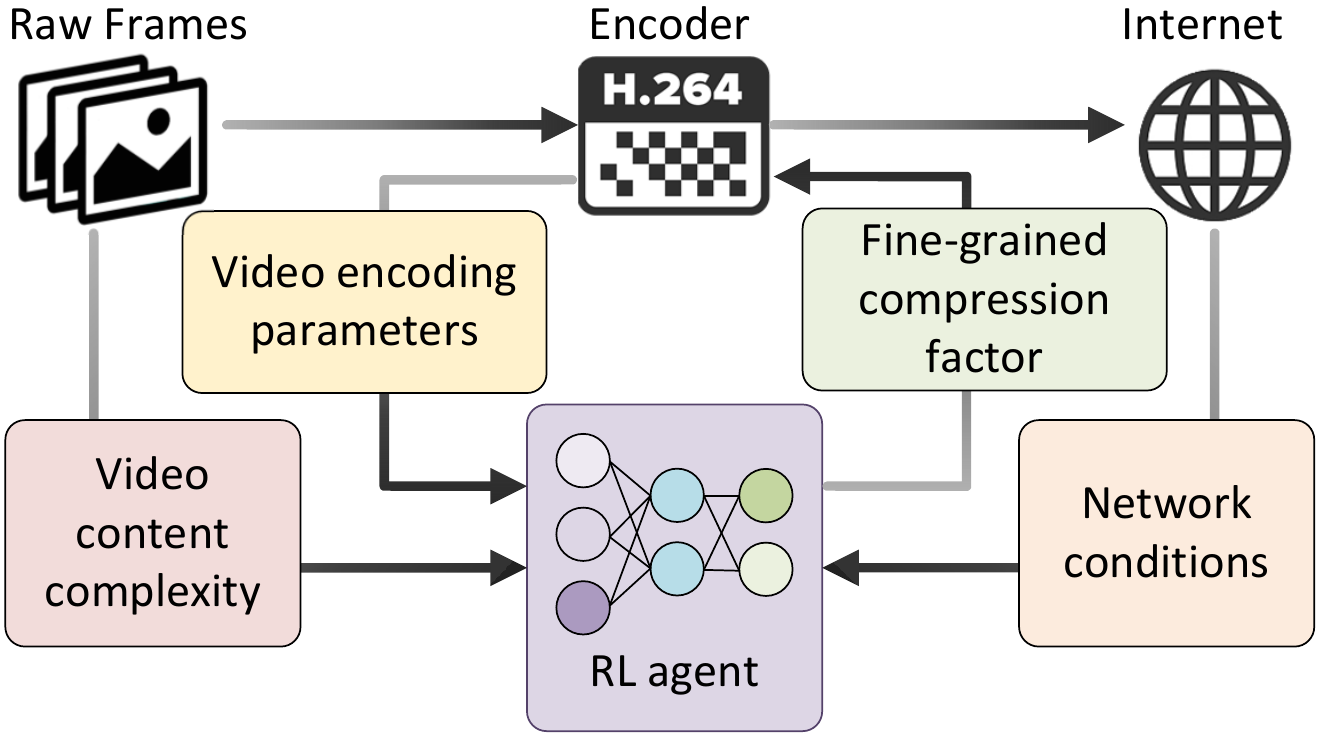}
    \caption{{\it Palette} leverages cross-layer factors to decide the fine-grained compression factor for optimal QoE. An RL agent is trained to capture the states and enforce the action recurrently. H.264/AVC compliant x264 is exemplified and other video codecs can be easily extended.}
    \label{fig:system_architecture}
\end{figure}

{\color{black} The architecture of Palette is shown in Fig.~\ref{fig:system_architecture}. First, Palette continuously collects past states of cross-layer factors from the environment, including the network conditions, video encoding parameters, and video content complexity (extracted from raw frames). Then Palette relies on the RL agent to map observed states to a fine-grained compression factor (e.g., average CRF for upcoming frames), which is then translated to per-frame QP to compress future video frames for network delivery. This decision will be passed to the codec and applied to encode future video frames until a new decision is made. We also designed a frame-level simulator to accelerate the training process, which faithfully models the process of a real-world end-to-end RTVC session.}

\subsection{QoE-oriented Cross-layer Factors}
\label{ssec:factors}

The proposed Palette specifically attempts to optimize the QoE directly, instead of the network utilization maximization as in existing works. It then motivates us to comprehensively define a set of cross-layer factors that are practically measurable and can be easily included in the RL engine for compression decision and QoE optimization. Referring to the discussions in Sec.~\ref{ssec:observations}, the QoE of networked video in RTVC systems can be modeled by the compressed video quality, the fluency of video playback and interaction delay, each of which is mainly impacted by one or more factors from the network, video encoder, and video content.

\textbf{Network conditions.} Network impairments would lead to video packet drops, severely deteriorating the video quality sensation at receiver~\cite{frame_stalling_quality,feng2010saliency}. We first include the RTT $d$ and packet loss rate $p$. Note that these factors are also used in other ABR solutions, and now can be easily measured in WebRTC clients. Additionally, fluent video playback at the receiver plays a vital role in QoE sensation~\cite{chen2019tgaming}. However, transport-layer factors like RTT and packet loss rate couldn't fully reflect playback status in the receiver client~\cite{Loki}. Thus, we also include the video stalling rate $h$, which is defined as the ratio of the number of stalled frames ($n^{'}$) to the total number of video frames ($n$) in a predefined time interval (0.2s as used here), e.g., ${h}={n}^{'}/n$. Although video stalling is used to reflect the playback status, we still categorize it into network factors because its occurrence is highly related to network behaviors.

\textbf{Video encoding parameters.}  Different from existing ABR approaches which simply model video quality by its bitrate, we use the encoder compression factor to infer the quality of the compressed video as suggested in~\cite{ou2010perceptual}. A common factor is the CRF $f$ (or equivalent QP) that determines the compression level and reconstruction quality of video frames. To exhaustively exploit the spatiotemporal correlation in the video, popular video encoders often adaptively apply the I-frame (or intra-frame) or P-frame (or inter-frame) encoding, mostly depending on the scene changes. Note that the I-frame can only explore spatial correlations while P-frame can utilize both spatial and temporal predictions, leading to different levels of compression efficiency. Even having the same CRF for I-frame and P-frame, they may present very different reconstructed qualities. Thus, the frame type $i$ is also used in our work.

\textbf{Video content complexity.} Following the discussion above, either I-frame or P-frame applies the predictive coding for redundancy removal with which its efficiency is highly dependent on the content complexity. Ideally, a relatively stationary content with a simple structure is easier to compress, otherwise, a motion-intensive frame with complex texture is typically hard to encode, largely yielding different perceptual sensations~\cite{ou2010perceptual}. And, encoding a video with complex scenes usually exhibits bitrate fluctuations along with the scene changes (e.g., bitrate jumps at around 2s and 13s as exemplified in the top part of Fig.~\ref{fig:contents}), which probably leads to bitrate overshooting induced stalling rate and delay increase.  


As also analytically deduced in~\cite{ou2010perceptual,ma2011modeling}, both perceptual quality and bitrate of a compressed video can be well modeled using the function of compression factors like CRF or QP, and video content complexity. Therefore, we suggest including the video content complexity in Palette to better adapt the bitrate and quality of the compressed video. For simplification, well-known SI $u$ and TI $v$ metrics standardized in ITU-R BT.1788~\cite{ITU-R_BT.1788} are used in this work as the representation of content complexity. We downsample raw video frames in both temporal and spatial dimensions for deriving the SI and TI in Palette, because of negligible difference from the SI/TI computation using non-downsampled video, and much less computational overhead.

\subsection{DRL Model}
\label{ssec:drl_model}
{\color{black} Palette learns to generate ABR policies using deep reinforcement learning, which has demonstrated superior performance in prior ABR works~\cite{chen2019tgaming, Pensieve}.} 
At each time step $t$,  the RL agent observes the state ${{s}_{t}}$ from the environment in a live RTVC session and takes an action ${{a}_{t}}$ according to its policy ${{\pi }_{\theta }}({{s}_{t}},{{a}_{t}})$ with the policy parameter $\theta$. After applying the action, the environment transits to a new state $s_{t+1}$ and returns a reward $r_t$ to the RL agent. This discretized interaction is conducted from one interval to another recurrently. In default, we set the interval to 0.2s, roughly 6 frames for a 30Hz video.  Here,  the ``time step'' can be connected with the ``interval'' to simplify the notation. Palette represents its policy using a neural network (NN) to fully explore its powerful representation capacity to capture the dynamic behaviors along with the recurrent process.
Following paragraphs detail the state, action, reward, and NN model used in Palette.

\textbf{State.} The state at $t$ is defined as  $s_t$=$(\vec{u_t},\vec{v_t},\vec{i_t},\vec{f_t},\vec{h_t},\vec{d_t},\vec{p_t})$, and each of them is a vector of values observed in past $k$ intervals (e.g., $\vec{u_t}=\{u_{t-k+1},\ldots,u_t\}$).
Here, for the same $t$-th interval, $u_t$ and $v_t$ represent the average SI and average TI of the raw frames within it; $i_t$ is a binary flag indicating the occurrence of the I-frame. If an I-frame is marked, the flag is 1; otherwise, it is 0. $f_t$ is the CRF value applied to all frames within $t$-th interval, and $h_t$, $d_t$ and $p_t$ are the stalling rates, average RTTs, and packet loss rates calculated in this interval, respectively. All of these are normalized to the range of [0,1] for model training. In this work, we set $k=6$ to capture dynamics from the past 6 intervals with a duration of 1.2s (6$\times$0.2s).

\textbf{Action.} The output of Palette's neural network is the probability distribution of actions. In order to ensure temporal smoothness, the change of CRF between adjacent intervals shall be as minimal as possible. Thus we set delta CRFs as the action set, i.e., $a_t\in{\{+8,+4,+2,0,-1,-2,-4\}}$. The RL agent chooses one of them according to the action probability, and then the video encoder immediately derives the CRF parameter. Currently, Palette collects states and performs actions every 0.2s. The impacts of other interval settings are discussed in Sec.~\ref{ssec:ablation_study}.

\textbf{Reward.} In an RL task, the agent learns the optimal policy by maximizing the expected cumulative (discounted) reward that it receives from the environment. Thus, we set the reward $r_t$ to reflect the QoE of the previous time interval in a live RTVC session. For RTVC services, the quality of the encoded video, the fluency of video playback, and video delay are three key QoE-related contributions. We then define the reward function in Palette as follows: 
\begin{equation}
\label{eq:reward_function}
    {{r}_{t}}=\lambda {q}_{t}-\mu {d}_{t}-\nu {h}_{t}.
\end{equation}
{\color{black} Here, ${{q}_{t}}$ is the difference between the encoding CRF value and the maximum CRF value (e.g., 51 in x264) at time step $t$,} which directly reflects the quality of the encoded video as extensively studied in~\cite{ou2010perceptual,ma2011modeling,crf}. ${{d}_{t}}$ refers to the average RTT (in ms), which is used to approximate the interaction delay between the sender and the receiver for RTVC services. And ${{h}_{t}}$ represents the stalling rate, which directly affects the fluency of video playback. The weighting parameters $\lambda$, $\mu$, and $\nu$ have a significant impact on policy learning. For example, a larger $\lambda$ leads to an ``aggressive'' policy pursuing high video quality, while a larger $\mu$ or $\nu$ results in a ``conservative'' policy that gives priority to ensuring the fluency of video playback. In Palette, these parameters are set to $10$, $0.12$, and $70$ empirically through extensive simulations. 

\begin{figure}[t]
    \centering
    \includegraphics[width=0.95\linewidth]{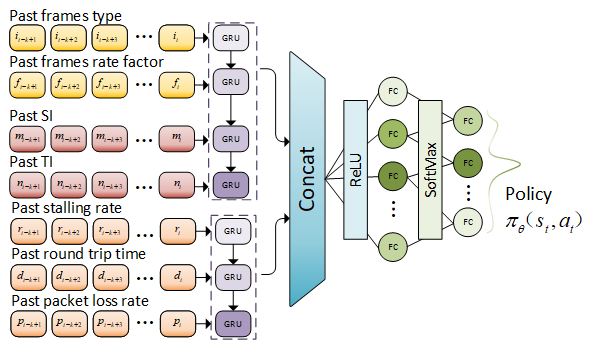}
    \caption{{\it Neural Model.} Palette applies the neural network (NN) to map observed states of cross-layer factors from the past to infer the action (e.g., CRF determination in this work).}
    \label{fig:NN_model}
\end{figure}

\textbf{Neural network architecture.} Palette uses the asynchronous advantage actor-critic (A3C~\cite{pmlr-v48-mniha16}) algorithm to learn its policy. And we design a lightweight neural network structure for Palette to extract features from the input states and process them for making decisions quickly and effectively. Fig.~\ref{fig:NN_model} shows the structure of the actor network. Specifically, two GRU layers each with 16 units are leveraged to extract video content/encoder features and network-related features from the input states, respectively. These features are then concatenated by a flatten layer, and passed into a fully connected network with 32 neurons. Both two layers use rectified linear units (ReLU) as the activation function. At the output layer, we utilize a fully connected network and SoftMax activation function to get a 7-dimension vector, which represents the probabilities of taking each action. As for the critic network, we use the same NN structure as the actor except for the output layer, which is designed as a linear neuron without activation function.

Note that we have tried another neural network with deeper layers, but experiments show that a deeper network does not bring significant performance improvement while greatly increasing the computational overhead.
On the other hand, we have also attempted to use a common GRU layer to extract features from all input states together, whereas an inferior performance is achieved compared to default settings in Palette in which separate GRUs are used for characterizing the behaviors of the network and video application layers. 

\subsection{Training Palette}
\label{ssec:training}
In this paper, we use the policy gradient method as the training strategy~\cite{NIPS1999_464d828b}. The main idea of the policy gradient method is to estimate the path of the fastest increase in the total return by observing the expected total reward, and update the network parameters according to this path. The gradient of the accumulated reward with respect to policy parameter $\theta $ can be written as follows~\cite{Pensieve,mao2020network}:
\begin{equation}
    {{\nabla }_{\theta }}{{E}_{{{\pi }_{\theta }}}}\left[{\sum\limits_{t=0}^{\infty }{{{\gamma }^{t}}{{r}_{t}}}}\right]={{E }_{{{\pi }_{\theta }}}}[{{\nabla }_{\theta }}\log {{\pi }_{\theta }}(a|s){{A }^{{{\pi }_{\theta }}}}(a|s)].
\end{equation}
Here ${{A }^{{{\pi }_{\theta }}}}(a|s)$ is the {\it advantage function}, which measures the advantage of a certain action $a$ compared to other actions given a certain state $s$. And we use ${A}({{a}_{t}}|{{s}_{t}})$ as an unbiased estimate of ${{A }^{{{\pi }_{\theta }}}}({{a}_{t}}|{{s}_{t}})$, which is empirically computed by the {\it value function} (denoted as ${V}({{s}_{t}};\theta_v)$) as follows:
\begin{equation}
\begin{aligned}
    {A}({{a}_{t}}|{{s}_{t}})=r_t+\gamma{V}({{s}_{t+1}};\theta_v)-{V}({{s}_{t}};\theta_v).
\end{aligned}
\end{equation}
The critic network learns an estimate of ${V}({{s}_{t}};\theta_v)$) from empirically observed rewards, and outputs estimated advantage to guide the training of the actor network. 
Thus, the update formula of the actor network parameters $\theta$ and the critic network parameters $\theta_v$ can be respectively represented using~\cite{pmlr-v48-mniha16}:
\begin{equation}
\begin{aligned}
\begin{split}
    d\theta \leftarrow d\theta &+\alpha \sum\limits_{t}{{{\nabla }_{\theta }}}\log {{\pi }_{\theta }}({{a}_{t}|{{s}_{t}}})A ({{a}_{t}|{{s}_{t}}})\\&+\beta {{\nabla }_{\theta }}H ({{\pi }_{\theta }}(\cdot |{{s}_{t}})),
\end{split}
\end{aligned}
\end{equation}
\begin{equation}
    d\theta_v \leftarrow d\theta_v - \alpha^{'} \sum\limits_{t}{{{\nabla }_{\theta_v }}}({\sum\limits_{i=0}^{t}{{{\gamma }^{t-i}}{{r}_{i}}}}-{V}({{s}_{t}};\theta_v))^2,
\end{equation}
where $\alpha$ and $\alpha^{'}$ are learning rates for actor and critic networks, respectively. Following~\cite{Pensieve,mao2020network}, we introduce the entropy of the policy (denoted as $H(\cdot)$) to avoid converging to sub-optimal policies at the early stage of training, and $\beta$ is the weight of the entropy term. In this work, we configure $\alpha=0.00025$, $\alpha^{'}=0.0015$, $\beta=0.15$, and $\gamma=0.9$. 
More details about the training can be found in \cite{pmlr-v48-mniha16,Pensieve}. Note that the training strategy is not the focus of this paper, and can be replaced by other methods.


To accelerate the training process, we design a trace-driven simulator, in which a video frame doesn't have to be actually streamed. In the simulator, we calculate the packet loss rate, RTT, and other network indicators just according to the time-aligned video bitrate and bandwidth samples. And we collect video content/encoder-related states based on the frame-level records of a video profile. To further accelerate the training, we deploy 16 work agents to explore the environment independently in parallel, and a central agent to aggregate all the experiences to generate a single ABR algorithm model. 

We train Palette on a server with a 16-thread Intel E5-2620 CPU and a Nvidia GTX-1080Ti GPU. And it costs us about 5 hours to complete the model training. We conduct the training multiple times, each of which reports a similar time cost, showing that Palette can converge to the optimal policy stably.

\section{Evaluation} \label{sec:evaluation}
We implement a WebRTC-based testbed - {\it Echo} to extensively and fairly conduct comparative studies with other prevalent ABR methods. Both lab-managed trace simulations and real-world field trials in the wild are exhaustively examined.

\subsection{Methodology}\label{ssec:eva_Methodology}

\begin{figure}[t]
    \centering
    \includegraphics[width=\linewidth]{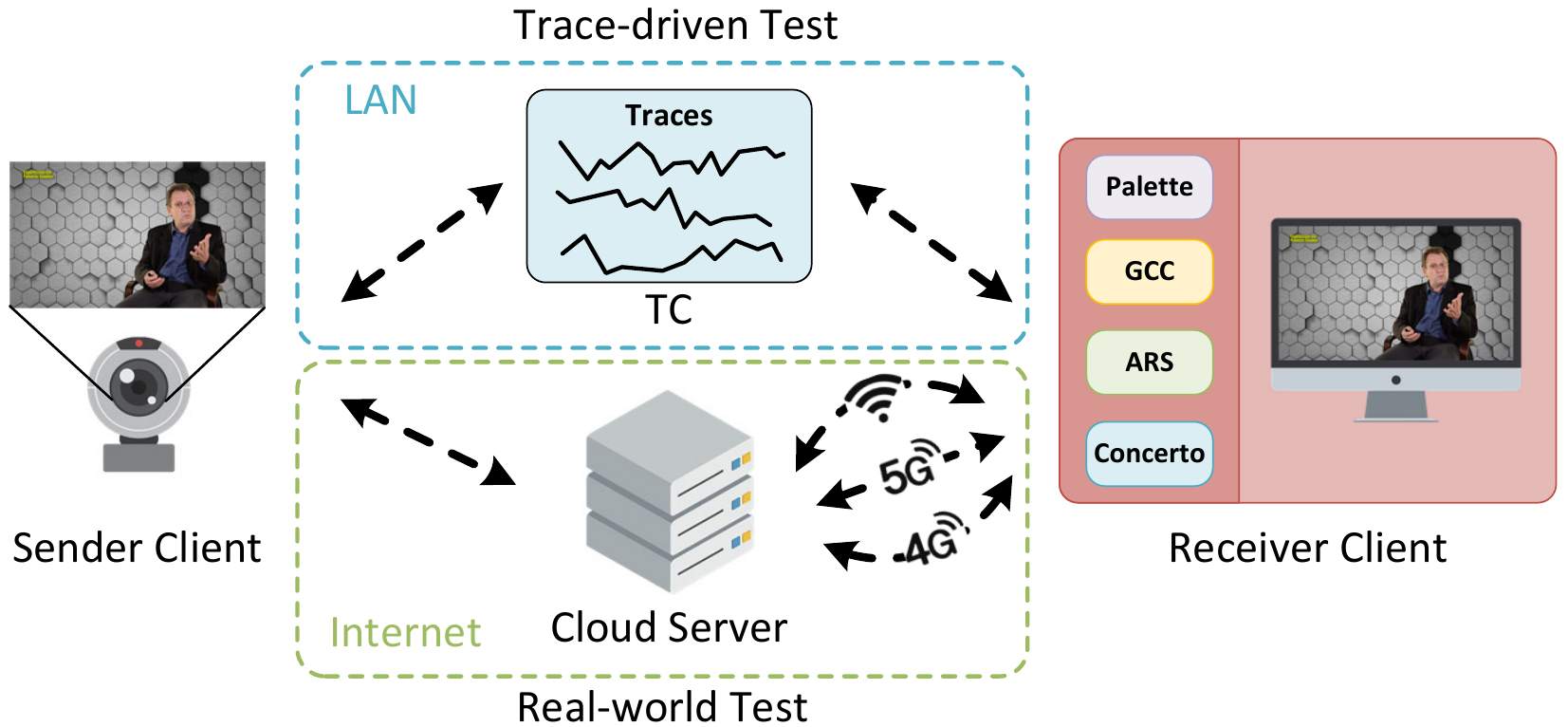}
    \caption{{\it Echo} is a real-time video conferencing testbed that complies with the WebRTC standard. Placing the server node differently allows us to perform the trace-driven simulations in a managed lab or directly run real-world field trials in the wild. Four ABR solutions are deployed for the study. {\it We will later make this testbed open to the public for reproducible research.}}
    \label{fig:Echo}
\end{figure}

\textbf{Testbed.} {\it Echo} is a WebRTC standard~\cite{webrtc_standard} compliant real-time video conferencing testbed shown in Fig.~\ref{fig:Echo}.  It is comprised of a sender client, a receiver client, and a server node in the middle to emulate or experience real-world network connections. {\it Echo} uses the default and most widely-adopted H.264/AVC~\cite{AVC_overview} compliant x264~\cite{x264} 
as the underlying video encoder. Other video codecs like VP8~\cite{VP8} can be easily enabled since our solution does not require any special customization on the video codec.

For trace-driven tests in the lab, we deploy two clients on different Windows PCs and the server on a Linux PC within the same LAN. We run the Linux TC tool on the server node to emulate network behaviors following the traces. For real-world tests in the wild, we rent a commercial cloud server from a leading Internet service vendor, and set the server-to-receiver down-link with 4G, 5G, or WiFi accesses in various environments (walking, bus/subway riding, caf\'e, etc.), for simulating a variety of real-world scenarios. Note that sender and receiver clients are deployed in different cities worldwide. {\color{black} We directly deployed the model trained on the training dataset in the receiver client without further training or tuning for different network scenarios.}



As discussed in the measurement study in Sec.~\ref{sec:measurement_setup}, we play pre-cached video files to simplify the simulations. The sender encodes live scenes in real time and then transmits the video stream to the receiver client. To emulate a variety of scenes, we randomly select video clips from YouTube UGC video dataset~\cite{8901772}, including gaming, television, animation, etc., to form a 1-minute long video, and then repetitively render it during a live RTVC session spanning about 10 minutes. To best mimic the real RTVC applications on the large-scale Internet, we configure to encode this long video at a relatively low resolution (i.e., 720P) to guarantee both the video quality and the playback fluency.

At the receiver client, we collect the RTT, packet loss rate measured by the WebRTC internals~\cite{webrtc} and proactively measure the stalling rate along with the video consumption. All four ABR methods are devised in the receiver client for evaluation. Note that we can deploy these ABR solutions on the sender side or intermediate cloud node as well~\cite{chen2019tgaming}.

\begin{figure*}[t]
    \centering
    \subfigure[HSDPA (${{b}_{avg}}=1.51$, ${{b}_{var}}=0.65$)]
    {
        \begin{minipage}[t]{.48\textwidth}
            \centering
            \includegraphics[width=\linewidth]{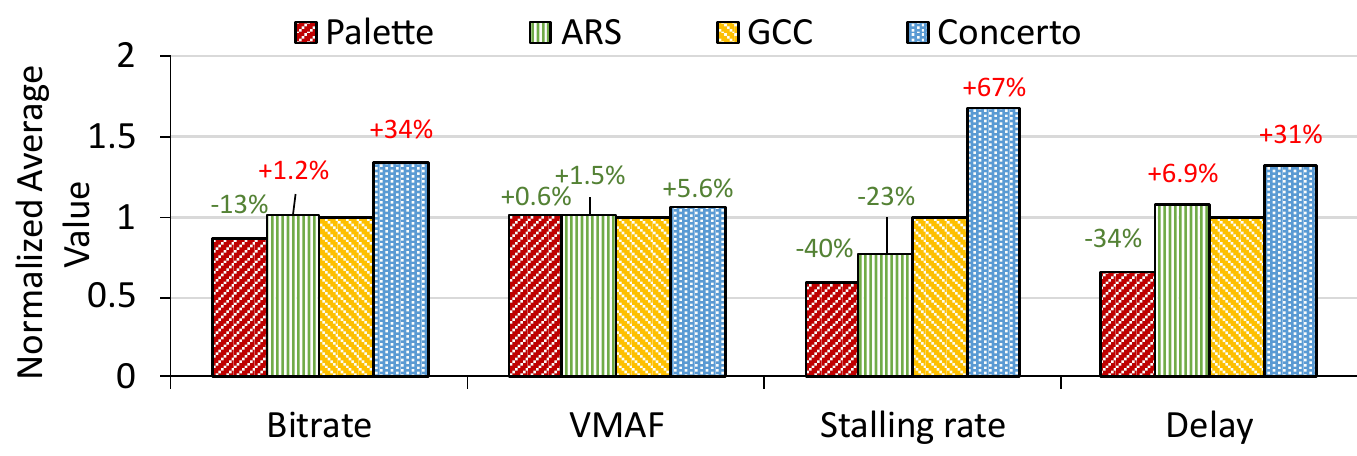}
            \label{fig:trace_hsdpa}
        \end{minipage}
    }
    \subfigure[Oboe (\bm{${{b}_{avg}}=2.65$}, ${{b}_{var}}=0.68$)]
    {
        \begin{minipage}[t]{.48\textwidth}
            \centering
            \includegraphics[width=\linewidth]{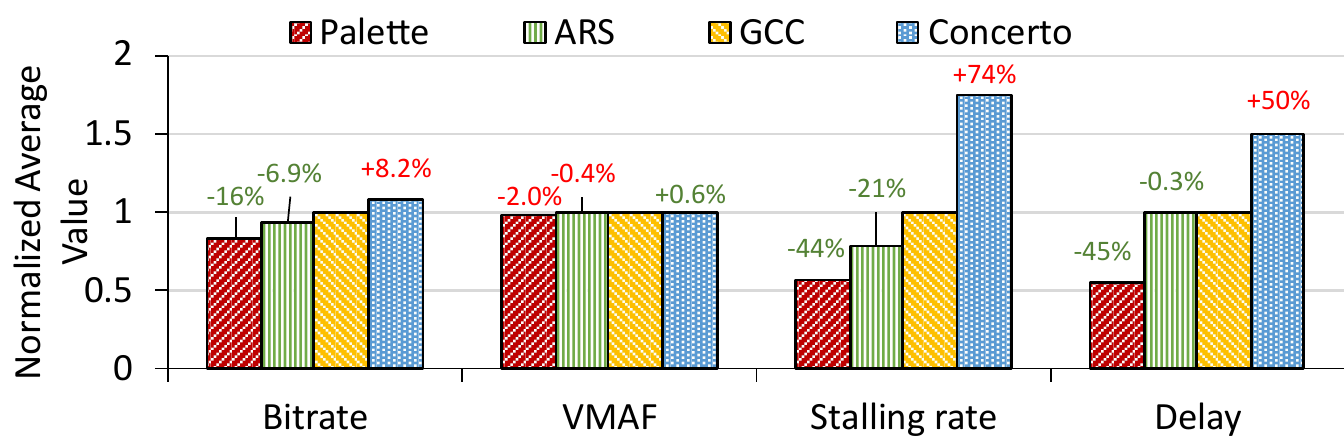}
            \label{fig:trace_oboe}
        \end{minipage}
    }
    
    \subfigure[FCC (${{b}_{avg}}=1.67$, ${{b}_{var}}=0.95$)]
    {
        \begin{minipage}[t]{.48\linewidth}
            \centering
            \includegraphics[width=\linewidth]{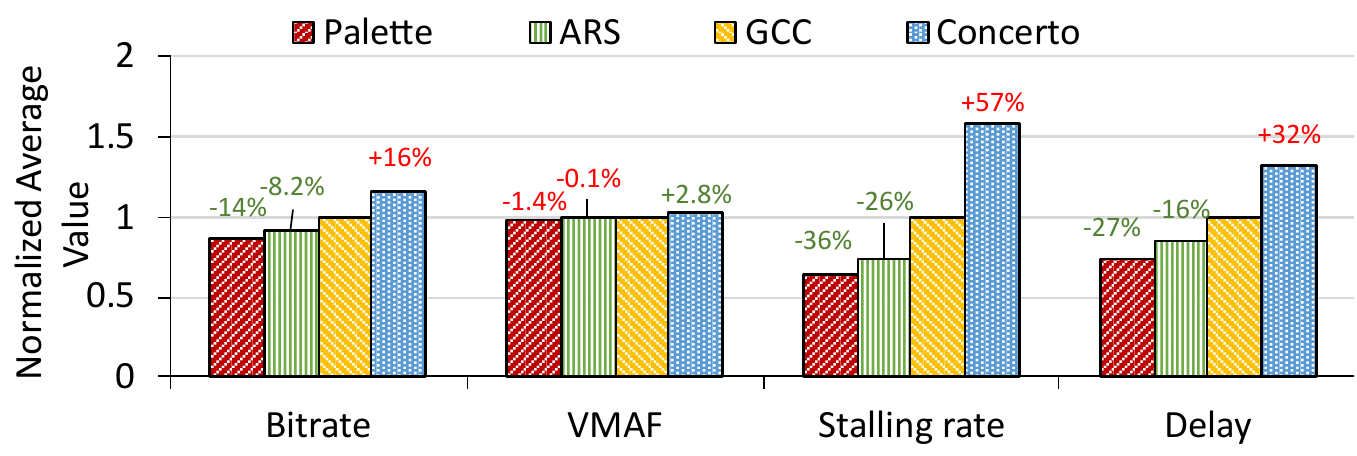}
            \label{fig:trace_fcc}
        \end{minipage}
    }
    \subfigure[MMGC2019 (${{b}_{avg}}=2.12$, \bm{${{b}_{var}}=1.61$})]
    {
        \begin{minipage}[t]{.48\linewidth}
            \centering
            \includegraphics[width=\linewidth]{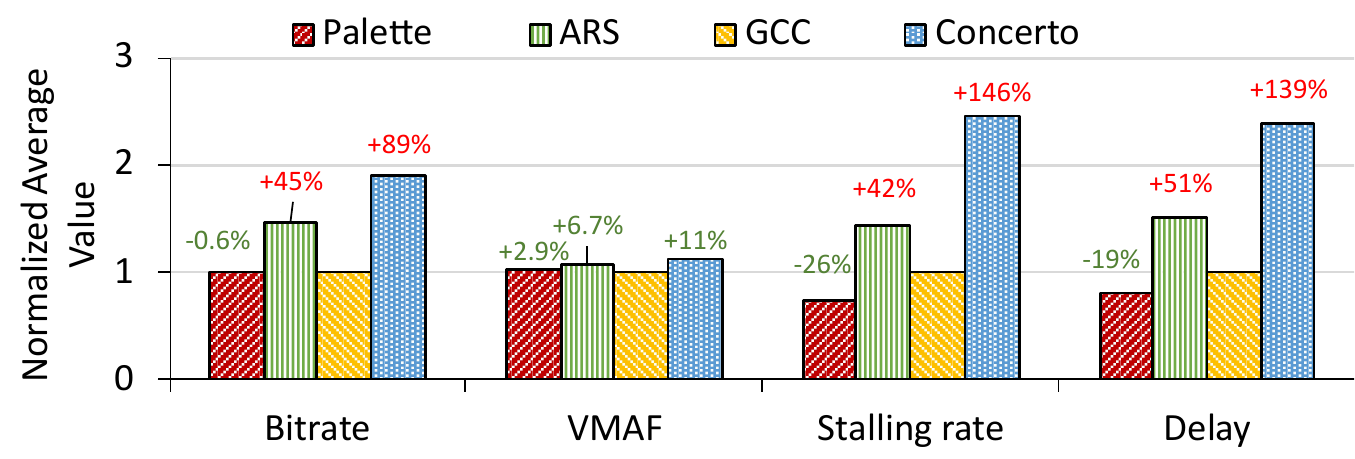}
            \label{fig:trace_mmgc}
        \end{minipage}
    }
    \caption{{\it Average Gains.} Metrics are averaged for all traces in the same category, and then normalized against the anchor using GCC as the ABR approach on {\it Echo}. {Average bandwidth $b_{avg}$ and bandwidth variation $b_{var}$, measured in Mbps, for each trace dataset are also provided.}    } 
    \label{fig:results}
\end{figure*}

\textbf{Network traces.} Similar to other works~\cite{10.1109/TNET.2017.2703615,chen2019tgaming,zhou2019learning}, we perform lab-managed simulations using well-known public network traces including: (1) HSDPA~\cite{10.1145/2483977.2483991}: a 3G/HSDPA network trace dataset collected using mobile devices when riding on vehicles; (2) Oboe~\cite{Oboe}: a Wi-Fi/cellular network trace dataset collected on both desktops and mobile devices {with relatively higher network bandwidth on average}; (3) FCC~\cite{fcc2016}: a broadband network trace dataset containing over 1 million traces, and (4) MMGC2019~\cite{mmgc2019}: traces collected 
for live video streaming competition {with much higher bandwidth variation to cover a wider range of scenarios.}
As seen, these traces are collected from a variety of access networks for different applications, best mimicking real-world network behaviors. We list the average and variance of network bandwidth of these traces in the subtitle of Fig.~\ref{fig:results}. 

It is worth noting that the network trace used for training has an impact on ABR policy learning. For example, when the network bandwidth maintains at a very high or extremely low level for a long period of time, selecting the best or worst video quality is always the optimal solution, which hinders the training from fastly converging. To avoid the occurrence of such situations, we filter out network traces whose minimum network bandwidth is below 0.5Mbps, and whose average throughput is higher than 4.0Mbps.
All traces in use also have a fine time granularity with network probing and bandwidth collection interval of less than 0.5s. Following the suggestions in previous works, we divide all traces into two parts randomly, and use 80\% of them for training and the rest 20\%  for testing. Overall, our training and testing sets include about 500 and 120 traces, respectively.

\textbf{State-of-the-art ABR approaches} for RTVC scenarios. 
\begin{itemize}
\item {\bf Google congestion control (GCC)}~\cite{10.1109/TNET.2017.2703615} is a prevalent rules-based ABR approach which first dynamically estimates the network bandwidth {based on end-to-end delay variations\footnote{We use the latest GCC implementation with trendline filter for study.} and packet loss rates, and then adapt the video bitrate according to the bandwidth estimation. It is integrated into the WebRTC protocol as the default bitrate adaptation algorithm. WebRTC has been a de facto standard supported by almost all major Web browsers for sustaining the RTVC services.} 
\item {\bf Adaptive real-time streaming (ARS)}~\cite{chen2019tgaming} is a RL based ABR approach. It trains a neural network to predict the encoding bitrate for the next group of pictures (GOP) according to the network and playback states observed from the past. Note that ARS was initially developed with a relatively high bitrate for cloud gaming, and this work scaled the bitrate range and retrained it under the same datasets as Palette and Concerto for fair comparison. 
\item {\bf Concerto}~\cite{zhou2019learning} is an IL-based approach that has extensively studied incoordination issues between the network transport and video codec. Though it suggested fusing the network bandwidth estimation and actual video bitrate from the past to get a better prediction, it still relied on the underlying video encoder to perform the rate control at the second scale, which, to the same extent, did not practically resolve the cross-layer incoordination. {\it As the Concerto is not an open-source project, we faithfully reproduced it following published technical details.}

\end{itemize}

\textbf{Evaluation metrics.} As the QoE measures the ultimate quality of video service at the receiver, it is affected by multiple factors from both network and video application layers. As a result, it is radically difficult to come up with a closed-form model for perfect quantification of QoE. Especially in a live RTVC session, a fast adaptation of network and video encoding would lead to very different sensations and of course, it is hard to model the QoE. Thus, we suggest using several popular metrics altogether, including the average video bitrate, average video quality, average delay, and average stalling rate, to have a more comprehensive evaluation. 

As for video quality, the CRF in Eq.\eqref{eq:reward_function} is not used since it is not available in other methods. Thus,  we then apply the VMAF~\cite{VMAF,vmaf_blog}, an objective perceptual quality metric widely acknowledged in the video quality assessment community, for a fair comparison.  Previous studies~\cite{Comyco} have reported that VMAF is better correlated with human perception than other metrics like video bitrate, structural similarity (SSIM)~\cite{SSIM}, etc. 
Both video bitrate and video quality are collected at the sender side to reflect the compression-induced QoE sensation.

Average RTT denotes the average delay over the entire session, and is defined as the sum of RTT sampled in each slot divided by the number of samples. The average stalling rate can be calculated by dividing the stalled video frames by the total video frames. Both RTT and stalling rate measure the potential QoE degradation impacted by the network which are augmented with sender-side VMAF and video bitrate to have an overall QoE sensation.



\begin{figure*}[t]
    \centering
    \includegraphics[width=\textwidth]{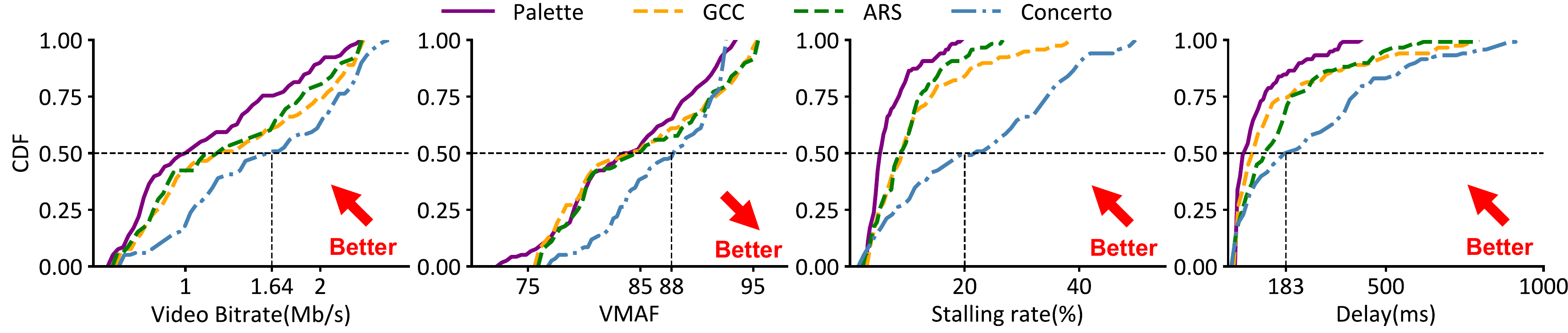}
    \caption{{\it Cumulative Gains.} Metrics are averaged for all traces, and presented in the form of CDFs.}
    \label{fig:resluts CDF}
\end{figure*}

\subsection{Trace-Driven Tests in Lab} \label{sec:trace_test}
We run {\it Echo} with four ABR approaches under different network characteristics using the aforementioned traces. Performance metrics, e.g.,  bitrate, VMAF, stalling rate, and delay (in RTT), are first averaged for each trace category, and then normalized against the anchor using the most popular GCC approach, as shown in Fig.~\ref{fig:results}. Besides,  the cumulative distribution functions (CDF) of those metrics are illustrated in Fig.~\ref{fig:resluts CDF}, having them averaged (but not normalized) across all traces.


{\bf Average gains.} As clearly reported in Fig.~\ref{fig:results}, the proposed Palette offers the greatest gains, consistently reducing  26\%-44\% of the stalling rate,  19\%-45\% of the delay, 0.6\%-16\% of the video bitrate, while maintaining almost the same video frame quality measured by the VMAF (i.e., variations within $\pm3$\% ), in comparison to the GCC anchor, across all trace sets. 

Performance gains are still retained for ARS although reductions of stalling rate and delay are less for most traces. Whereas, the Concerto severely increases them, when compared with the GCC anchor. This is because the stalling rate is also included in ARS for optimization, while the Concerto simply tunes the bitrate to maximize the bandwidth utilization. As seen, the video bitrate produced by the Concerto is increased, which may improve the video quality (e.g., up to 11\% VMAF gains against the GCC anchor in Fig.~\ref{fig:trace_mmgc}), but greatly increases the possibility of bitrate overshooting (see Fig.~\ref{fig:algs}) induced stalling and high delay.                                                                                                                                                                     
We particularly notice that the gains of the proposed Palette are the least for the MMGC2019 traces set in Fig.~\ref{fig:trace_mmgc}, and in the meantime, notable performance losses are also reported for both ARS and Concerto (e.g., respective $\approx$2.5$\times$ and $\approx$2.4$\times$ stalling and delay against the GCC anchor). After deeply analyzing the trace categories, we have found that this is mainly because the characterization of network behaviors is getting more and more difficult when the bandwidth variation enlarges gradually. As comparing the results for HSDPA in Fig.~\ref{fig:trace_hsdpa} and that for MMGC2019 in Fig.~\ref{fig:trace_mmgc}, the smaller network bandwidth variation comes with better performance, and vise versa. 

{\bf  Cumulative gains.} In addition to the average gains discussed in preceding sections, we again show that the proposed Palette reports clear and constant performance lead spanning the entire range, to other ABR approaches, for stalling rate (26.1\%-65.3\% decrease) and delay (39.5\%-55.1\% decrease) measurements. We notice that the VMAF CDFs are almost overlapped for Palette, GCC, and ARS, i.e., almost the same video quality at the sender, but Palette desires less bitrate (12.3\%-15.2\% saving), resulting in better rate-distortion performance~\cite{720552}. Similar to the aforementioned findings, the Concerto provides better VMAF than the GCC, ARS, and Palette at the sender side, but at the cost of larger compressed video bitrate, which may exceed the bandwidth, e.g., bitrate overshooting in both Fig.~\ref{fig:algs} and  Fig.~\ref{fig:contents}, and largely increase the stalling rate and delay. For illustration, we draw a horizontal line when $CDF=0.5$. Even though the VMAF is 88 for Concerto, the delay reaches 183ms and stalling rate reaches 20\%, implying that the compressed video frame may not possibly arrive at the receiver client due to a highly congested network.

{Interestingly, the delay and the staling rate are bounded with an upper limit in Palette as shown in Fig.~\ref{fig:resluts CDF}. However, other methods present a very long tail. Such a ``long tail'' problem is specifically visited in Loki~\cite{Loki} by fusing the rules-based and learning-based ABR approaches. Since this work mainly deals with the QoE-oriented ABR system development, it devotes the comparative studies to the GCC, ARS, and Concerto instead. Having the long tail problem partially resolved in Palette is an extra reward, revealing that Palette can best adapt itself to network fluctuations for uncompromised overall QoE. Improving Palette to thoroughly tackle the long tail problem as in Loki~\cite{Loki} is out of the scope of this work, and deferred as our future study.}

\begin{figure*}[ht]
    \centering
    \subfigcapskip = 2pt
    \subfigure[Overall]
    {
        \label{sfig:field_overall_results}
        \begin{minipage}[ht]{.305\linewidth}
        \centering
            \includegraphics[width=\linewidth]{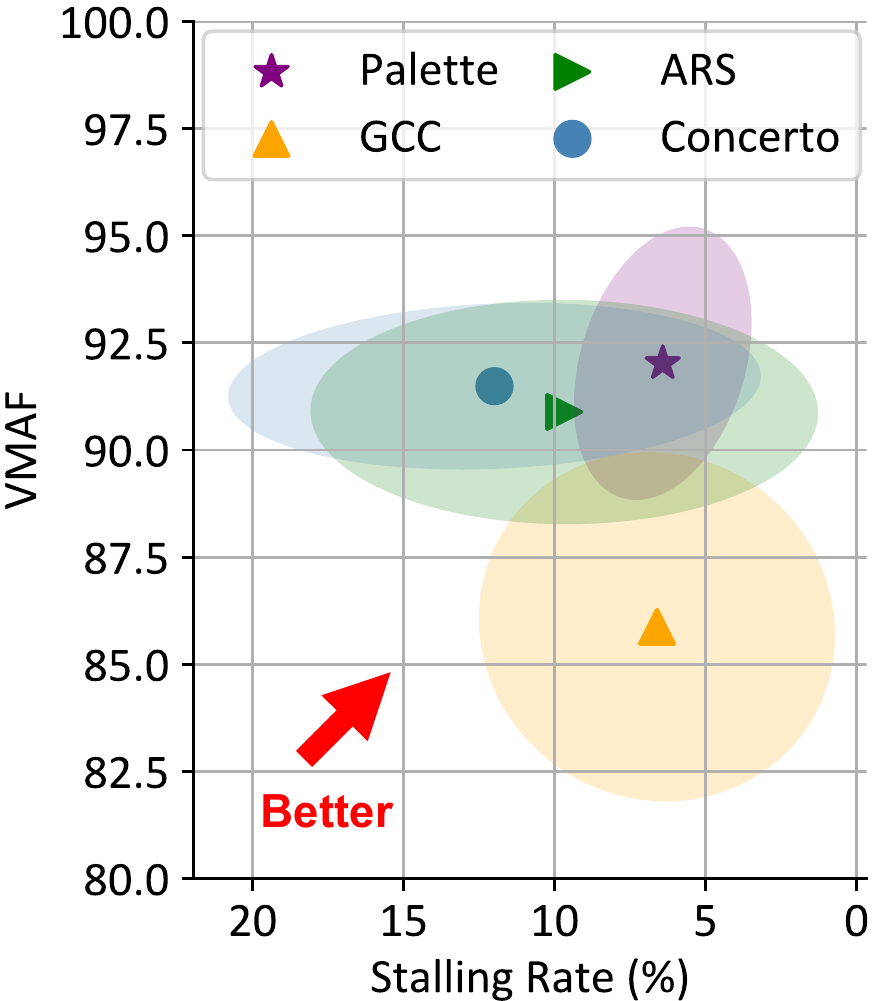}
        \end{minipage}
    }\hspace{-4pt}
    \subfigure[Various Scenarios w/ Diverse Access Networks and Scenes]
    {
        \label{sfig:field_consistent_results}
        \begin{minipage}[ht]{.665\linewidth}
            \includegraphics[width=0.32\linewidth]{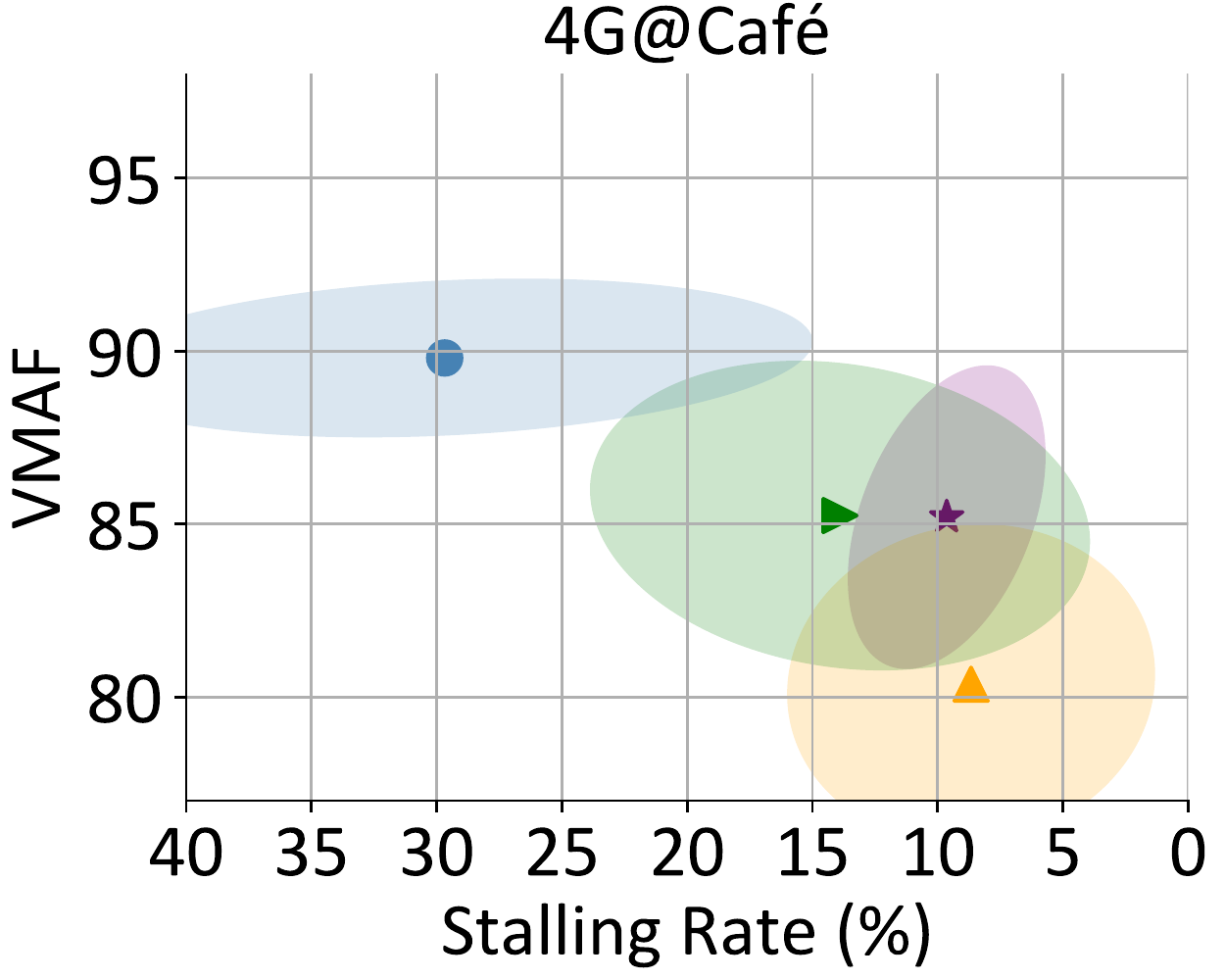}
            \includegraphics[width=0.32\linewidth]{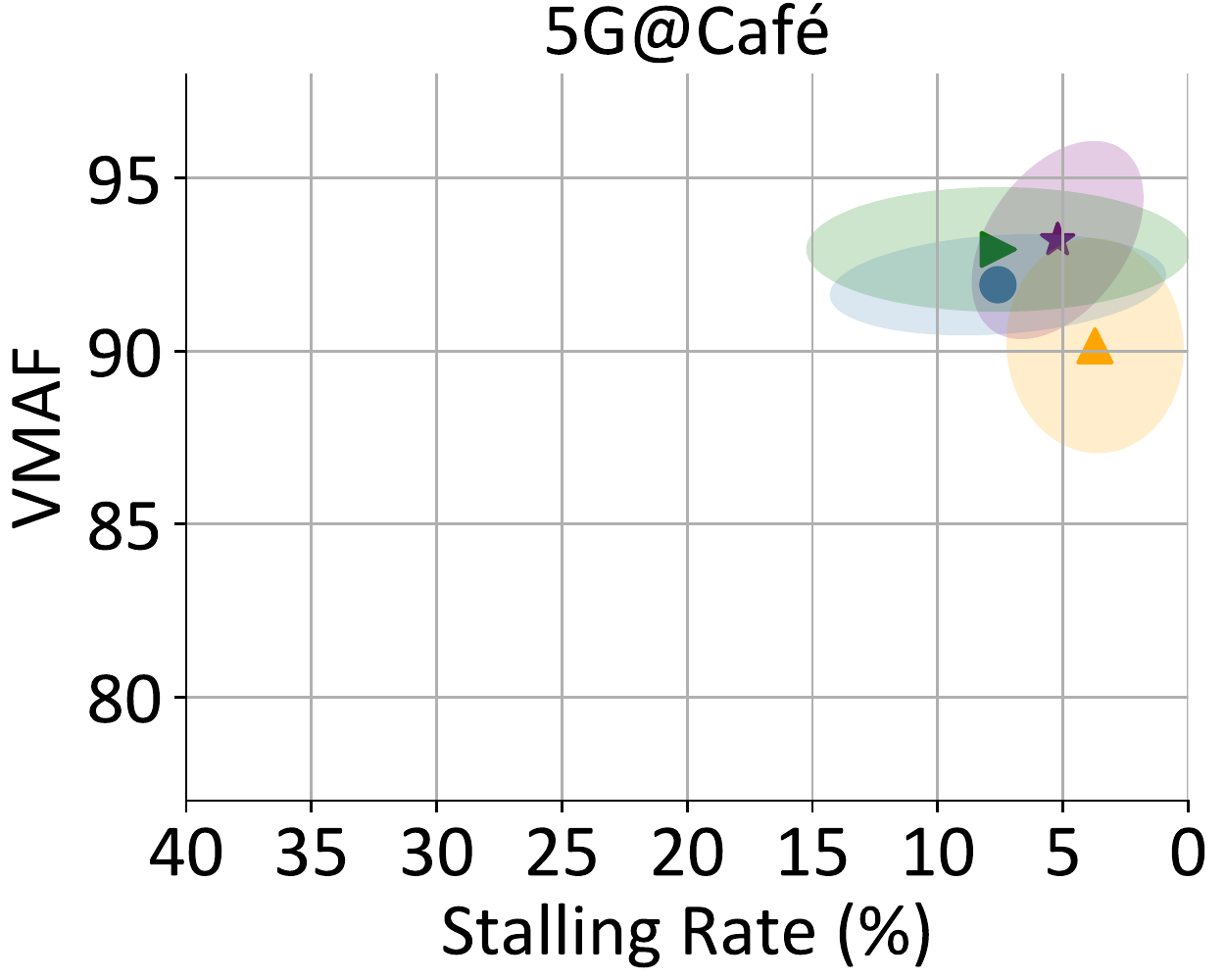}
            \includegraphics[width=0.32\linewidth]{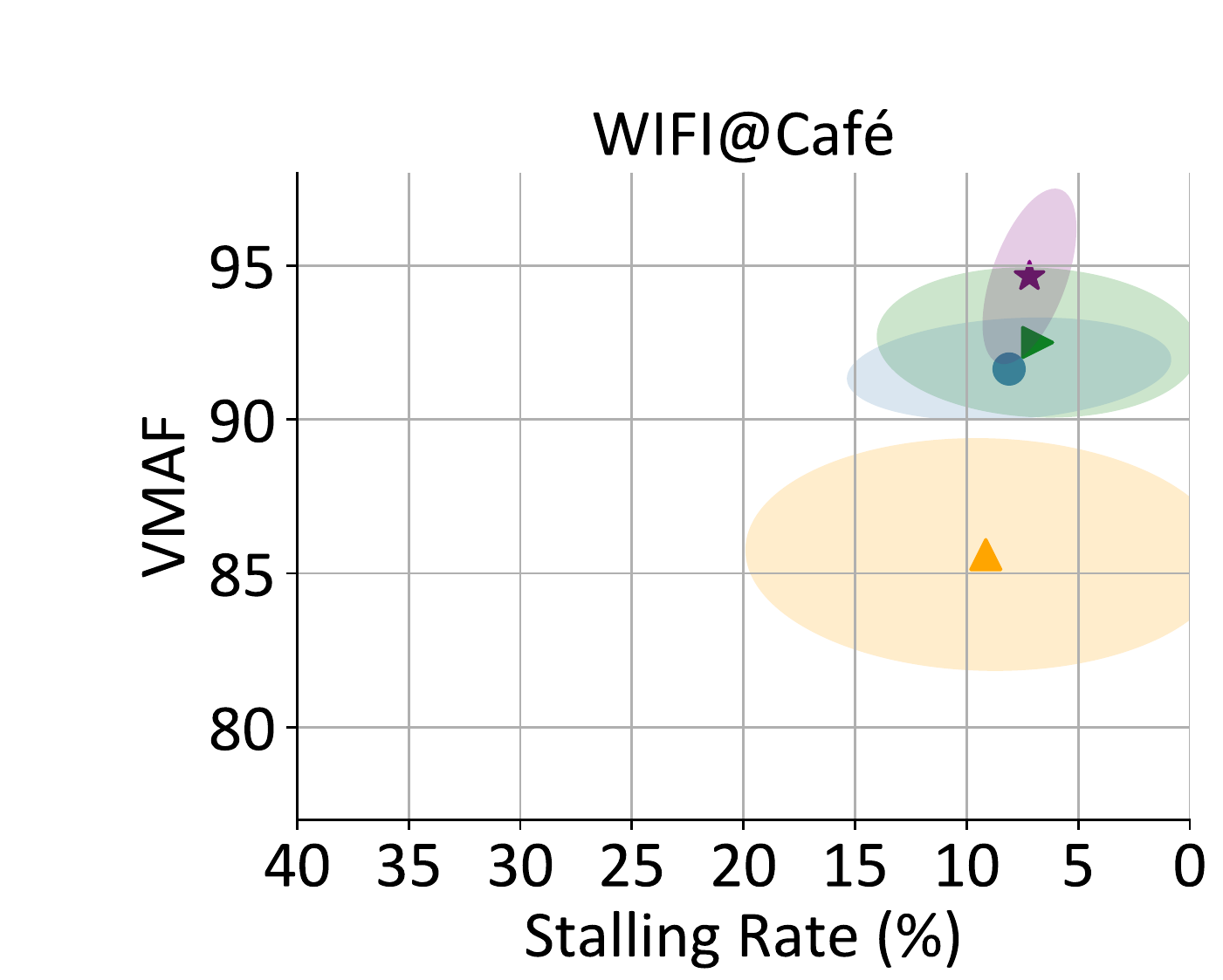}\vspace{-8pt}\\
            \includegraphics[width=0.32\linewidth]{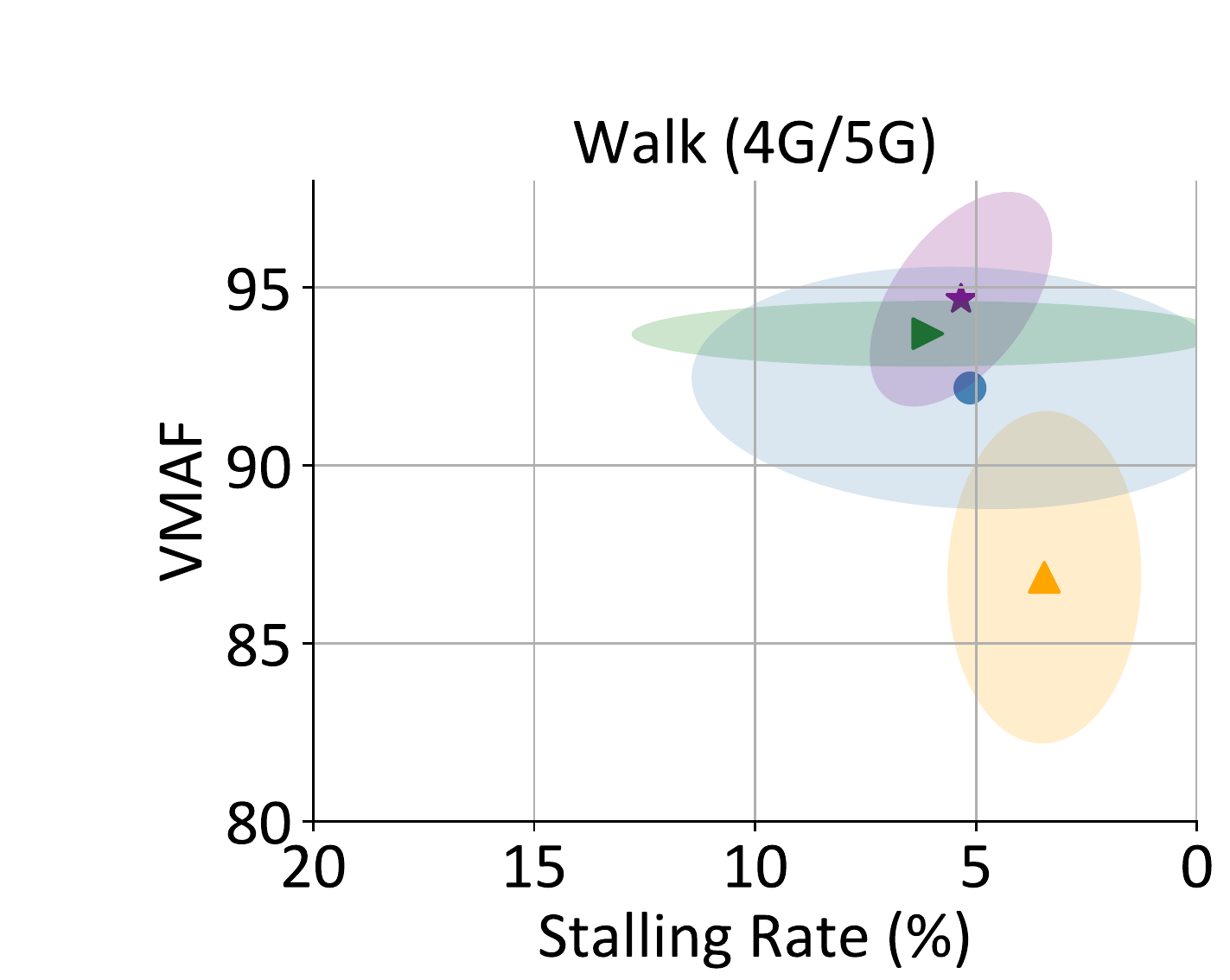}
            \includegraphics[width=0.32\linewidth]{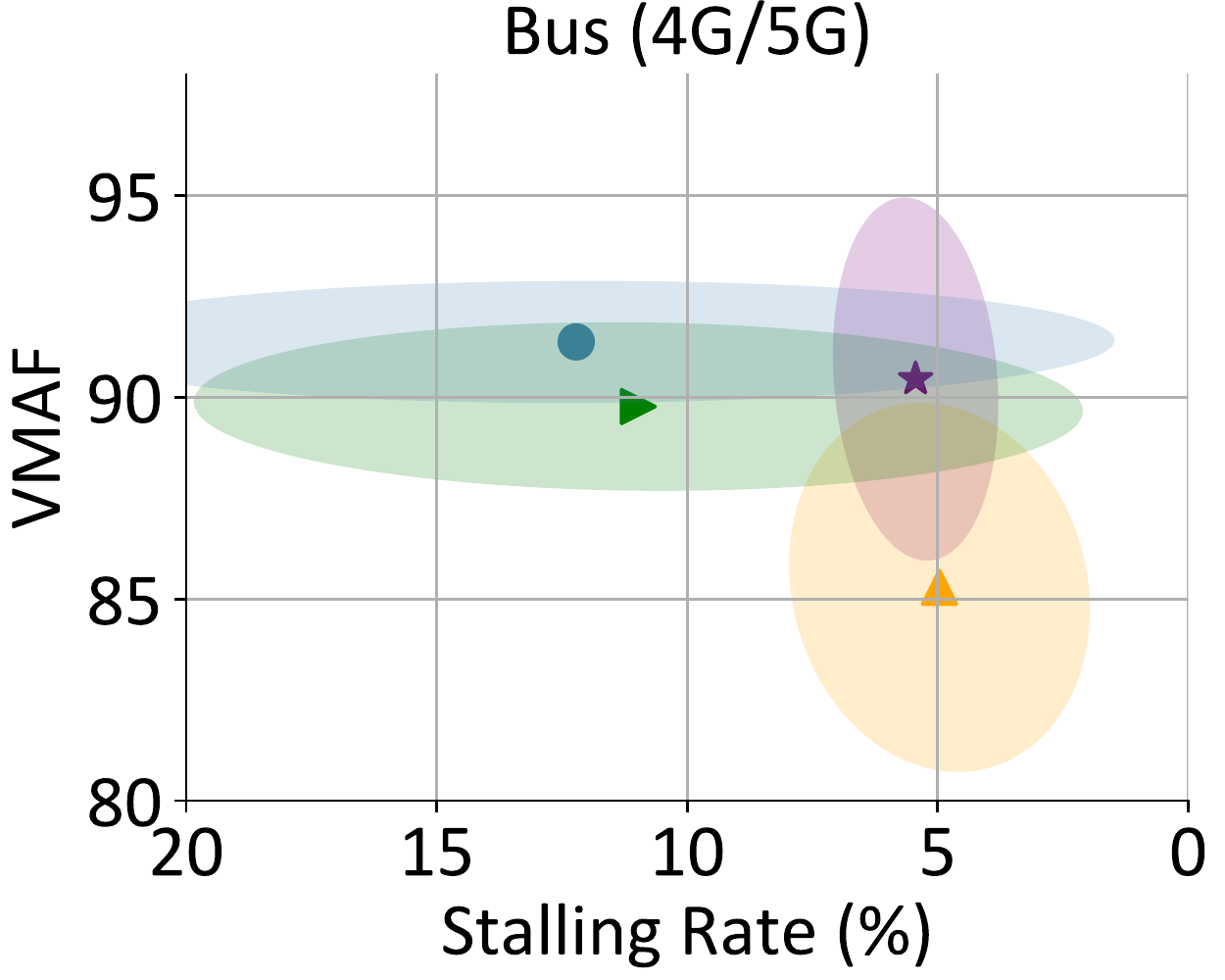}
            \includegraphics[width=0.32\linewidth]{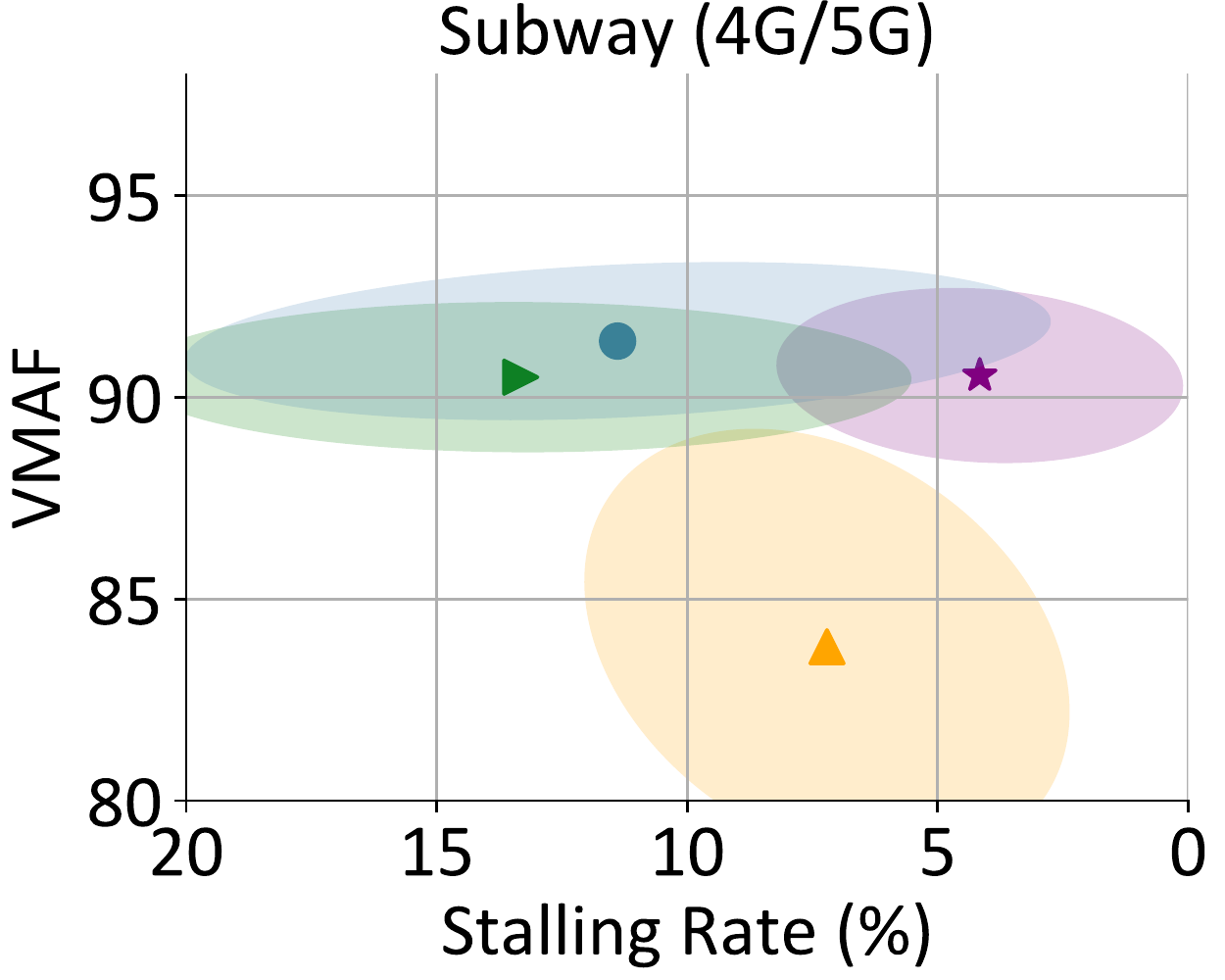}
        \end{minipage}
    }
    \caption{{\it Real-world Field Tests.} Palette shows consistent performance advantages to others across various access networks and scenes. The center of each ellipse shows the average values of paired VMAF and stalling rate, while the range of the ellipse in $x$- or $y-$ axis represents the standard deviation of  stalling rate or VMAF correspondingly.}
    \label{fig:field_resluts}
\end{figure*}

\subsection{Real-World Field Tests} 
Following the setup in Sec.~\ref{ssec:eva_Methodology}, we conduct real-world field tests with mainstream access networks, including the 4G, 5G, and WiFi under various scenes, e.g., caf\'e, walking, and bus/subway riding that are experienced day by day. Besides hosting the server node by a 3rd-party Internet cloud service vendor, {\it Echo} receivers are located in different cities hundreds of miles away in the same continent or even thousands of miles away across the continents.
{\color{black} We repeated an RTVC session 3 times for each algorithm in each scenario to ensure that the experimental environment was fair for all algorithms. Moreover, we made sure that all algorithms were connected to the same service provider under the same network type (4G/5G/WiFi) and experienced the same movement route and speed in the same mobility scene (Walk/Bus/Subway).} Each session lasts 10 minutes to fully experience a variety of video content for emulating numerous applications (e.g., gaming, conferencing, etc.). {And the entire field tests span several weeks.}

{\bf Overall evaluation.} Fig.~\ref{sfig:field_overall_results} presents the overall performance averaged across all sessions. As the encoder-side VMAF can be approximated monotonically using a logarithmic function of the bitrate~\cite{VMAF}, and the changes of stalling rate and delay share almost the same trends as in Fig.~\ref{fig:results}, we only plot the VMAF and stalling rate for simplicity. Their average and standard deviation are both used for four ABR methods to not only understand their performance on average, but also the performance variations induced by heterogeneous connections in terms of geographical locations, networks, scenes, etc. Overall, the proposed Palette offers a clear performance lead for all scenes and connections,
e.g., higher VMAF (0.2\%-7.2\% increase), less stalling rate (3.1\%-46.3\% reduction), and lower delay (20.2\%-50.8\% reduction) on average (eclipse center) as well as the smaller standard deviations of VMAF and stalling rate (e.g., smaller coverage of the eclipse). {Note that the overall performances for all approaches here are far better than that in the trace-driven test. This is because the real-world field tests are performed under a variety of high-bandwidth network connections (e.g., 5G and WiFi), which were filtered out in the trace-driven test.}


{{\bf Per-Scene evaluation.} Per-scene evaluation is detailed in Fig.~\ref{sfig:field_consistent_results}. Three subplots in the upper part of Fig.~\ref{sfig:field_consistent_results} exemplify the Caf\'e scene under 4G, 5G, and WiFi accesses, with which we wish to best represent stationary scenes at workspace, home, etc. Similarly three subplots in the bottom part of Fig.~\ref{sfig:field_consistent_results} show popular motion scenes, e.g., walking, and bus/subway riding, with more convenient 4G/5G connections, that are extensively experienced in our daily life.}

The stalling rate, including its average and standard deviation, for both Concerto and ARS is greatly increased in scenes like 4G@Caf\'e, Bus (4G/5G), and Subway (4G/5G). This is mainly due to the frequent network fluctuations in these scenes that cannot be effectively tackled by Concerto and ARS. For example, in  4G@Caf\'e, the number of users connected to the same 4G mobile radio tower 
varies dynamically from time to time. Because of the very limited bandwidth offered by 4G technologies as compared with the WiFi and 5G accesses, user dynamics would largely lead to notable bandwidth fluctuations. On the other hand, in scenes of  Bus (4G/5G) and Subway (4G/5G), network bandwidth usually fluctuates due to the change of radio tower along with the vehicle driving. By contrast, the proposed Palette can well capture the network fluctuation with better QoE (much less stalling rate and similar VMAF).

{
In the 4G@Caf\'e scene, although the Concerto shows the lead in VMAF, it significantly enlarges the stalling rate in terms of both average and standard deviation. This matches the observation in Sec.~\ref{sec:trace_test} where aggressively tuning the video bitrate to maximize the network utilization in Concerto can give better VMAF, but clearly increases the stalling rate. Again, having a better VMAF on the sender side may not guarantee the same level of QoE since network impairments would severely deteriorate the service.}
{Since much larger bandwidth can be offered by 5G and WiFi solutions than 4G networks, all ABR approaches in these scenes, e.g., 5G@Caf\'e and WiFi@Caf\'e, can offer better performances with higher VMAF and less stalling rate.}


 
{\bf Delay jitter.} Surprisingly, the GCC in real-world tests behaves differently from in trace-driven simulations.
As seen, it conservatively prefers a lower video bitrate with a relatively lower stalling rate but much degraded VMAF. This is because GCC is very sensitive to the network delay (RTT) jitter, which is widespread in the public Internet but not well emulated in throughput-based traces.

\begin{figure*}[t]
    \centering
    \begin{minipage}[t]{.3\linewidth}
        \centering
        \includegraphics[width=\linewidth]{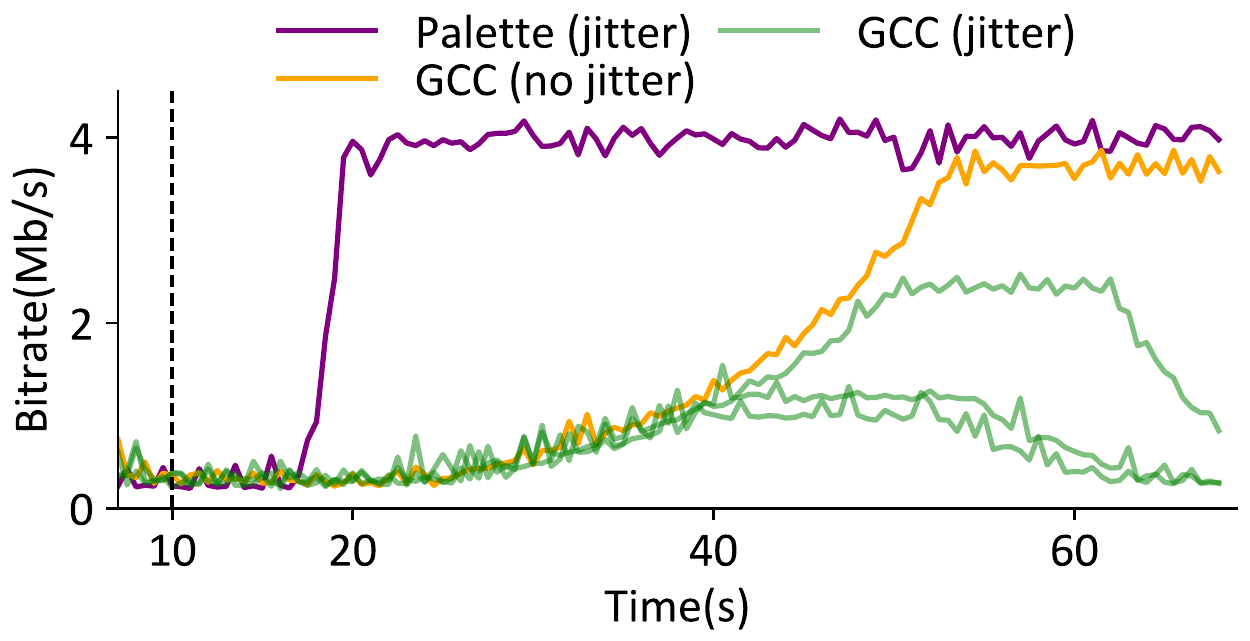}
        \caption{Delay Jitter Impact.}
        \label{fig:gcc_jitter}
    \end{minipage}
    \vspace{-2pt}
    \begin{minipage}[t]{.38\linewidth}
        \centering
        \includegraphics[width=\linewidth]{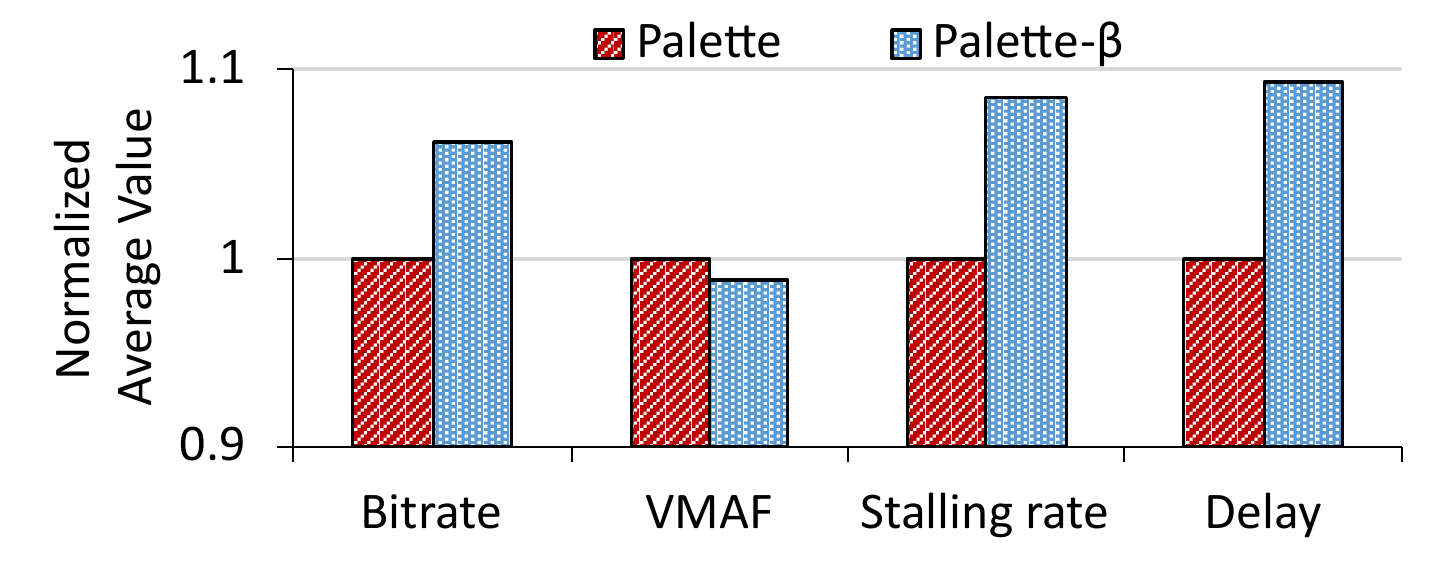}
        \caption{Palette w/o SI \& TI.}
        \label{fig:nositi}
    \end{minipage}
    \vspace{-2pt}
    \begin{minipage}[t]{.3\linewidth}
        \centering
        \includegraphics[width=\linewidth]{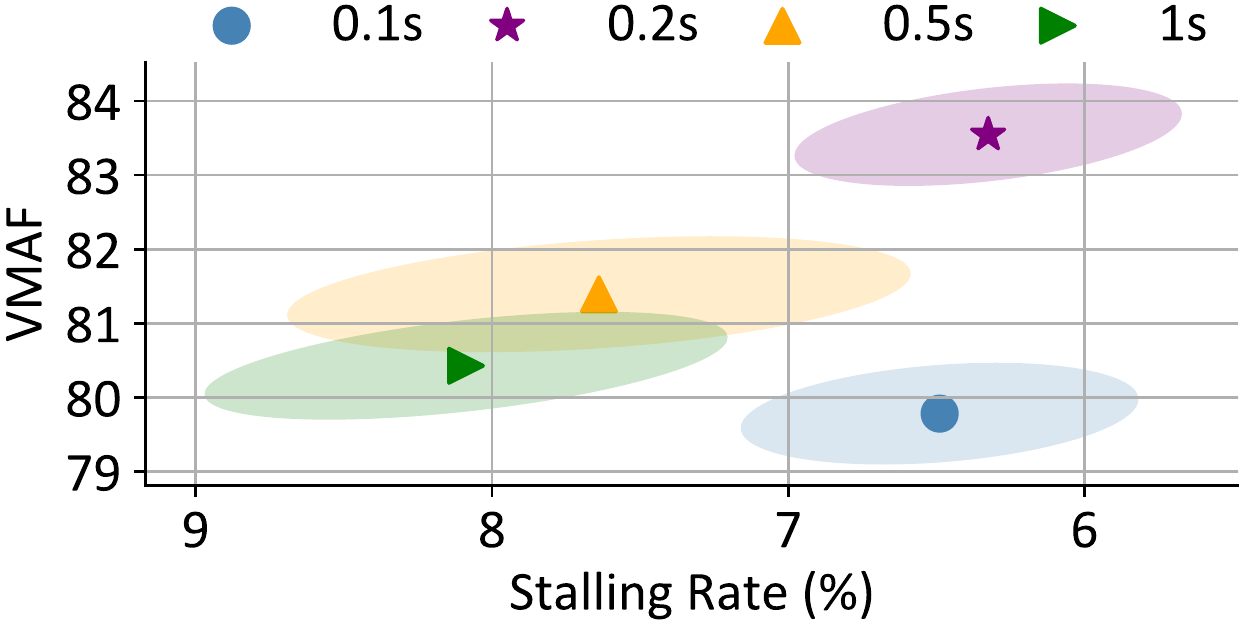}
        \caption{CRF Granularity.}
        \label{fig:granularity}
    \end{minipage}
\end{figure*}

To get more insights, we experiment with two similar network connections. In one connection (``no jitter''), the bandwidth and delay are fixed at 0.4Mbps and 100ms respectively for the first 10 seconds (before the dashed line in Fig.~\ref{fig:gcc_jitter}), and then they are respectively changed to 4Mbps and 90ms. The other connection (``jitter'') shares the same setting except for the delay varying in the range of $90ms\pm2ms$ after 10 seconds. Even having such a small delay variation, we already reproduce the GCC behavior as in real-world tests.


We then test GCC with the aforementioned two different settings, i.e., one with delay jitters and the other without jitter, respectively, and plot the video bitrate over time in Fig.~\ref{fig:gcc_jitter}. For the network connection without jitter (orange line), GCC gradually improves the estimation of available bandwidth, which leads to the increase of video bitrate until the match of the target at around 55s. By contrast, when having network jitter (green lines), GCC typically improves the bandwidth estimation first and then decreases it after a certain period, incurring similar occurrences of video bitrate (e.g., gradually increasing for a while and then dropping). We run the simulations multiple times, and almost the same trend is observed. This is mainly contributed by the conservative behavior of GCC when having network jitter.
Nevertheless, Palette (purple line) is able to recover and sustain the video bitrate close to the bandwidth eventually, regardless of network jitter.

\begin{figure}
    \centering
    \begin{minipage}[t]{\linewidth}
        \centering
        \includegraphics[width=\linewidth]{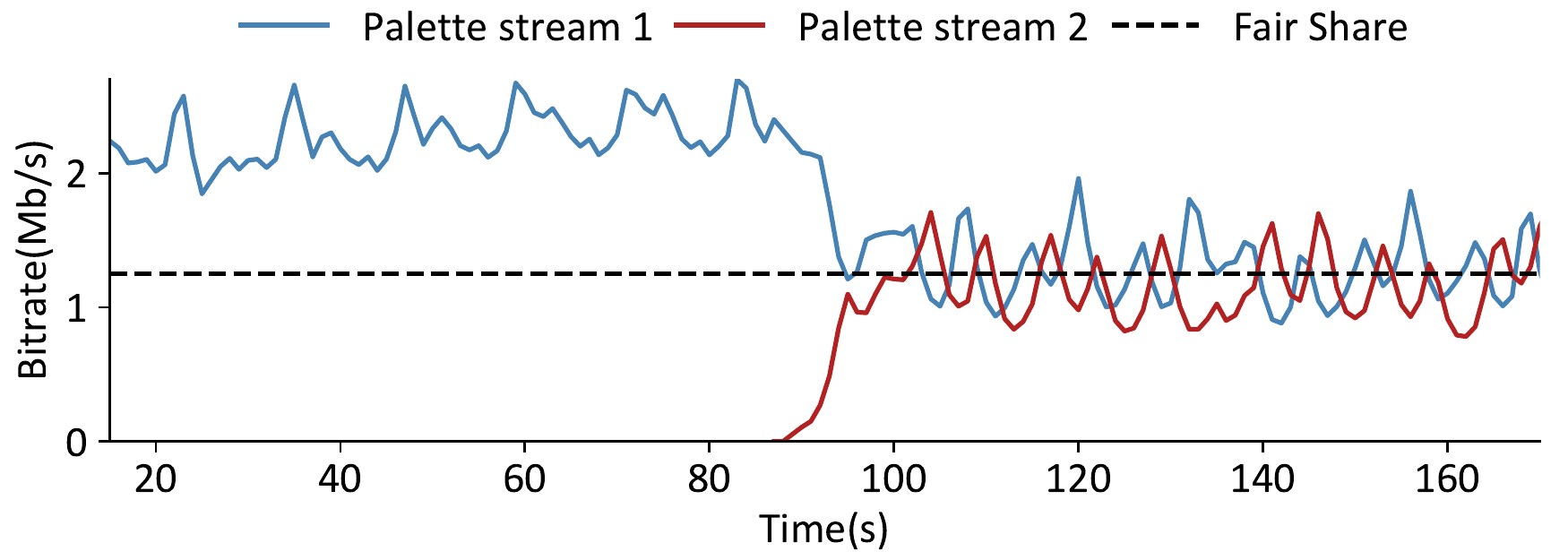}
        \caption{Competition between two Palette streams.}
        \label{fig:pltvsplt}
    \end{minipage}
\end{figure}

\subsection{Deeply Diving Into Palette} \label{ssec:ablation_study} 
This section dissects the Palette to get more insights.

\textbf{Content complexity.} Both SI and TI are introduced in Palette to reflect the contribution of video content on rate/quality control. Here we remove these content factors but only rely on network states and video encoding parameters for rate control which is noted as Palette-$\upbeta$. {Other settings remain the same for Palette-$\upbeta$ and the native Palette in both training and testing. We compare these two versions of Palette by running them under the same traces randomly selected from the datasets discussed in Sec.~\ref{ssec:eva_Methodology}.} As shown in Fig.~\ref{fig:nositi}, Palette-$\upbeta$ provides 8.6\% and 9.4\% increase on average respectively for the stalling rate and delay. In addition, although Palette-$\upbeta$ uses a larger bitrate, it still suffers a slight drop in VMAF. This is because Palette-$\upbeta$ wastes more bits across complex, high-motion frames, due to its inferior capacity to capture content characteristics without the embedding of SI/TI for accurate rate and QoE modeling in compression.




\textbf{Adaptation granularity of CRF.} The proposed Palette enables per-frame compression adaptation, where we often first determine an average CRF for a group of successive frames, and then let the encoder to further determine the QP for each frame. Such multiscale compression control allows us to better characterize the spatiotemporal correlations across frames for better compression factor decisions towards optimal QoE. 
Previous discussions assume the adaptation granularity (AG) of CRF at 0.2s.
Here we further explore other AGs at 0.1s, 0.5s, and 1s. As illustrated in Fig.~\ref{fig:granularity}, the default Palette with 0.2s-AG offers superior performance to other AG settings. For example, Palette with other AGs degrades the VMAF and increases the stalling rate, which to some extent, clearly impairs the final QoE. {This is because a smaller AG, e.g., 0.1s-AG, can only use state observations in past 0.6s, which are insufficient to make an optimal decision. Although larger AGs, e.g., 0.5s-AG and 1s-AG, can use abundant states, they   suffer from the ``adaption lag''  as discussed in Sec.~\ref{ssec:observations} (i.e., network behavior already varies in such a long duration).}

{\color{black}
\textbf{Multiple flows.} To investigate Palette's performance in a multi-flow competition scenario, we conducted experiments using the testbed described in Sec.~\ref{ssec:eva_Methodology}, which includes two pairs of senders and receivers responsible for transmitting two RTVC streams, both of which pass through the same relay server. We run the TC tool on the server to limit the bandwidth to a constant 2.5Mbps. The second stream joined around 90 seconds after the first stream. Fig.~\ref{fig:pltvsplt} shows the sending bitrate in competition between two Palette streams. We can see that when the second stream joins, the sending bandwidth of stream 1 quickly drops, mainly due to the congestion caused by the competition. After about 15 seconds, both streams reach and fluctuate around the fair bandwidth. The measured Jain Fairness Index~\cite{jain1984quantitative} reaches 0.993, indicating that both streams share the link bandwidth fairly.

}

\section{Discussion}

{\bf Generalization of Palette.} The Palette leverages the cross-layer states (e.g., network conditions, video encoding parameters, video content complexity) to derive the compress factor for the optimal QoE.  In the current implementation, Palette adapts the CRF in x264 for video compression and uses the A3C algorithm to train the policy for the mapping between cross-layer states and CRF. Whereas, Palette can be easily extended to support other video encoders, e.g., VP8~\cite{VP8}, x265~\cite{sullivan2012overview}, AV1~\cite{AV1}. And other advanced reinforcement learning algorithms, e.g., PPO~\cite{PPO}, DDPG~\cite{lillicrap2015continuous}, SAC~\cite{SAC}, and TD3~\cite{TD3}, can be applied to train Palette as well. 

{\bf Rate control of networked video.} Our extensive studies have reported that using cross-layer states for compression factor determination in Palette provides better rate control than the legacy approach embedded in the video encoder (e.g., as exemplified in GCC, ARS, and Concerto) when facing network bandwidth changes. This is because, 1) although legacy rate control in video encoder also uses content complexity across temporal reference frames for better rate estimation, it can only use a very limited number of references defined in standard specifications. While on the contrary, the proposed Palette can collect video content complexity from a large number of consecutive frames for better rate modeling. 2)  On the other hand, Palette learns a nonlinear mapping between content complexity and compression factor, which is clearly outperforming the simple quadratic or even linear function used in the video encoder. This work sheds light on the novel perspective to develop better video encoding rate control for networked video services.

{\color{black}
{\bf Video quality metric.} In Palette, we use CRF as the quality metric in the state space, action space, and reward, because it can obtain better video quality assessment without introducing any additional computational complexity compared to the previous methods which use bitrate as the video quality metric. We acknowledge that there are some full-reference metrics such as VMAF may better reflect subjective visual quality compared to CRF. However, due to that full-reference metrics require a copy of the original image which is not feasible in a live session on the client side, VMAF can not be used in the observed states of Palette. 

On the other hand, considering the nature of reinforcement learning, merely replacing CRF with VMAF in the reward function may not effectively improve the performance of Palette and could even lower training efficiency. Advanced reinforcement learning algorithms, including the A3C used by Palette, follow the actor-critic architecture. In this architecture, the critic's role is to fit the value function, which intuitively is a function mapping from the observed states to the reward. To fit this function, the neural network needs sufficient input information. Otherwise, the information gap between the input and output is just noise to the agent, which further reduces training efficiency. For example, VMAF computation involves a large number of pixel-level calculations, and predicting VMAF requires sufficient pixel-level information and a more complex information extraction network. Previous works, such as QARC~\cite{QARC}, designed a dedicated and complicated neural network to predict VMAF directly from past lossless video frames. In contrast, Palette's current design is based on a balance between image quality and computational complexity.}


\section{Conclusion}

We presented and evaluated Palette, a QoE-oriented ABR system that leverages cross-layer factors from both network and video application layers to decide fine-grained compression factors toward the optimal QoE in RTVC applications. 
Both trace-driven simulations in the lab and real-world tests in the wild picture that the proposed Palette outshines state-of-the-art approaches over a broad set of networks, scenes, and video contents.
We believe that Palette will inspire more ideas on cross-layer ABR research.

\bibliographystyle{IEEEtran}
\bibliography{reference}

\begin{IEEEbiography}[{\includegraphics[width=1in,height=1.5in,clip,keepaspectratio]{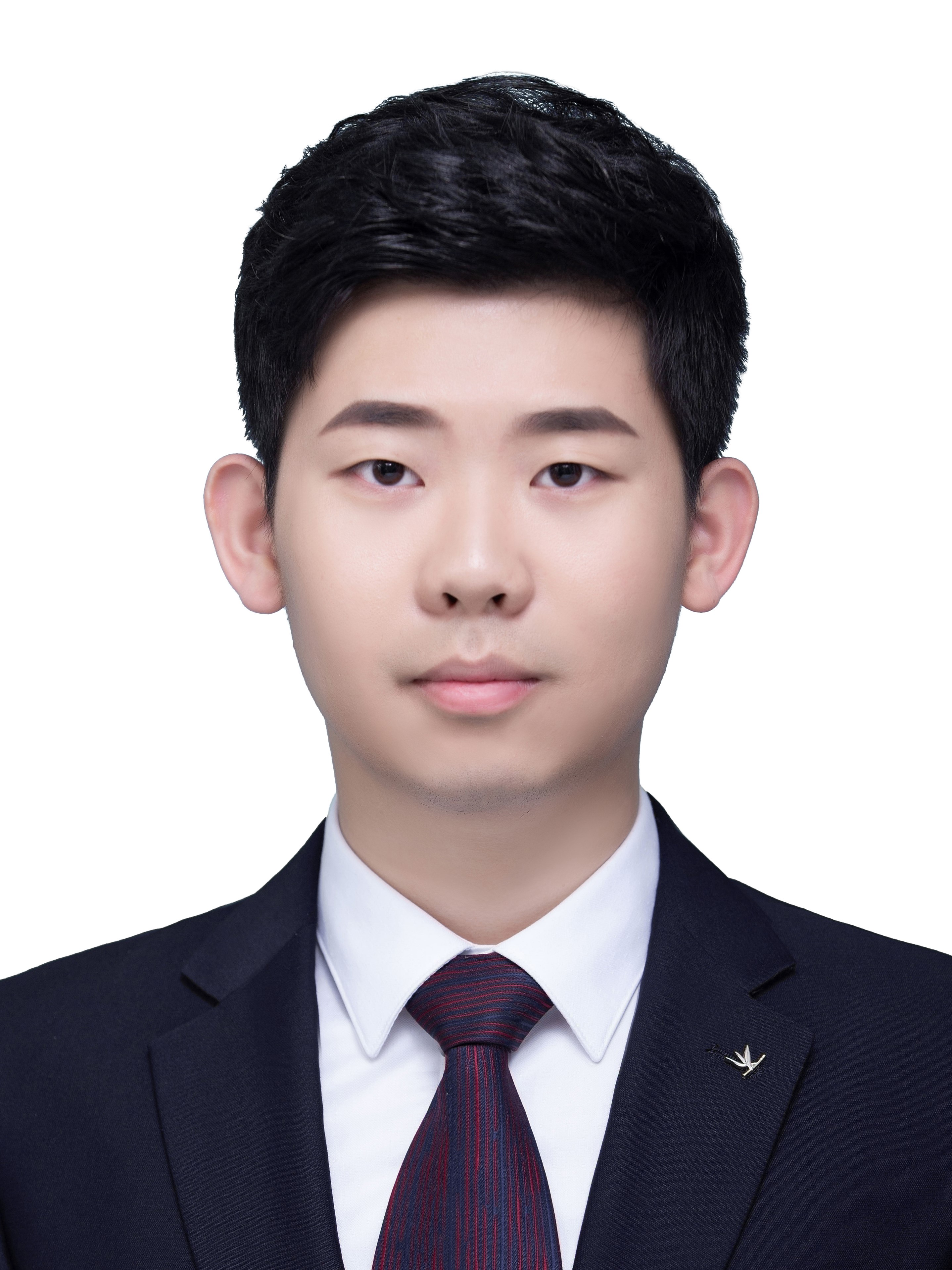}}]{Yueheng Li}
received the B.E. degree from the College of Electronic and Information Engineering, Nanjing University of Aeronautics and Astronautics, in 2021. He is currently pursuing the M.S. degree with the School of Electronic Science and Engineering, Nanjing University. His research interests include live video streaming and real-time video communication.
\end{IEEEbiography}

\begin{IEEEbiography}[{\includegraphics[width=1in,height=1.5in,clip,keepaspectratio]{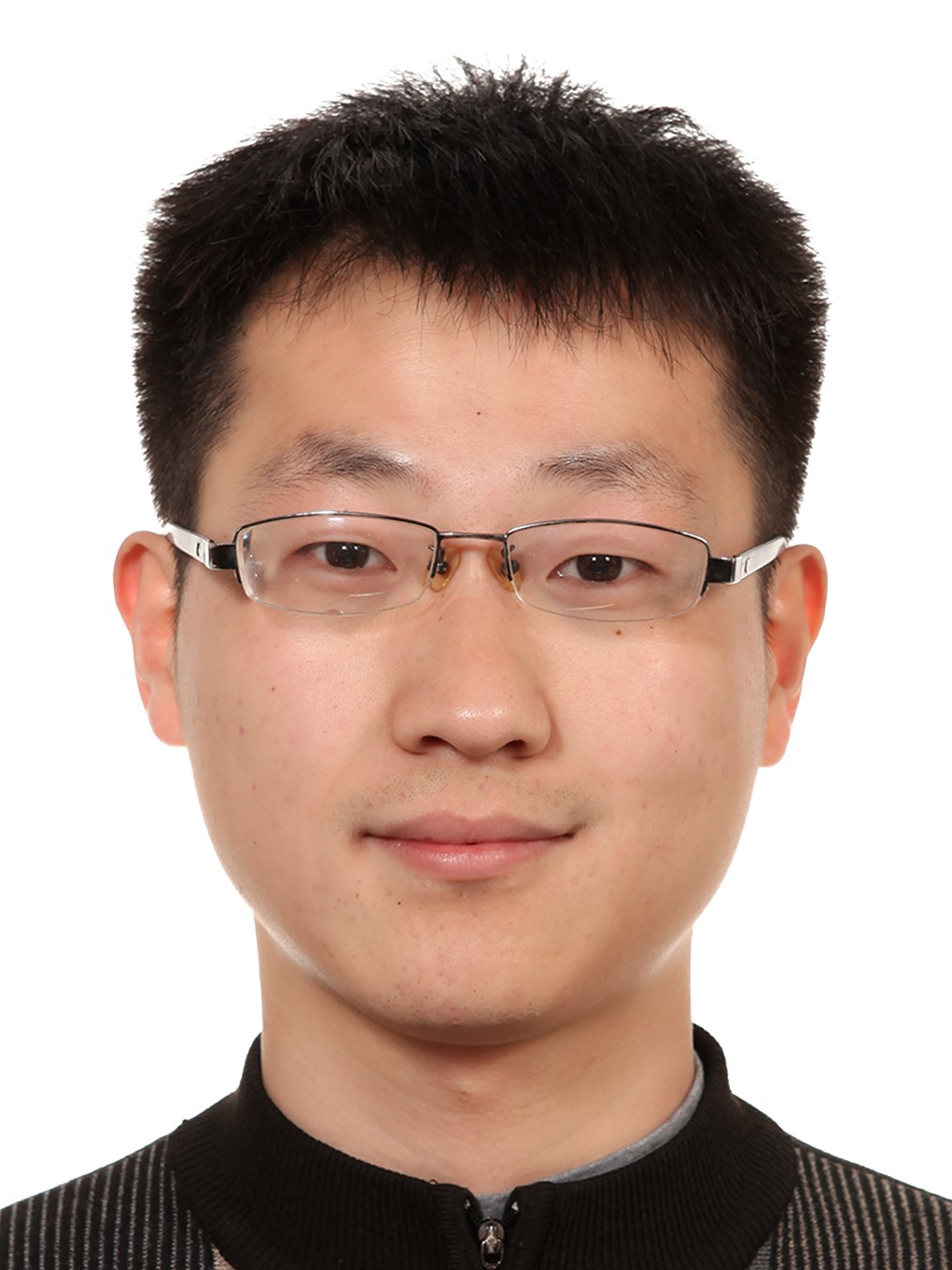}}]{Hao Chen}
(Member, IEEE) received the B.E. degree in electronics and information engineering from Northwestern Polytechnical University, China, in 2013, and the PhD degree from Shanghai Jiao Tong University, China, in 2020. He is now on the faculty of Electronic Science and Engineering School, Nanjing University. His research focus on video streaming, real-time video trasnmisison and machine learning. He is a co-recipient of 2019 IEEE Broadcast Technology Society Best Paper Award.
\end{IEEEbiography}


\begin{IEEEbiography}[{\includegraphics[width=1in,height=1.5in,clip,keepaspectratio]{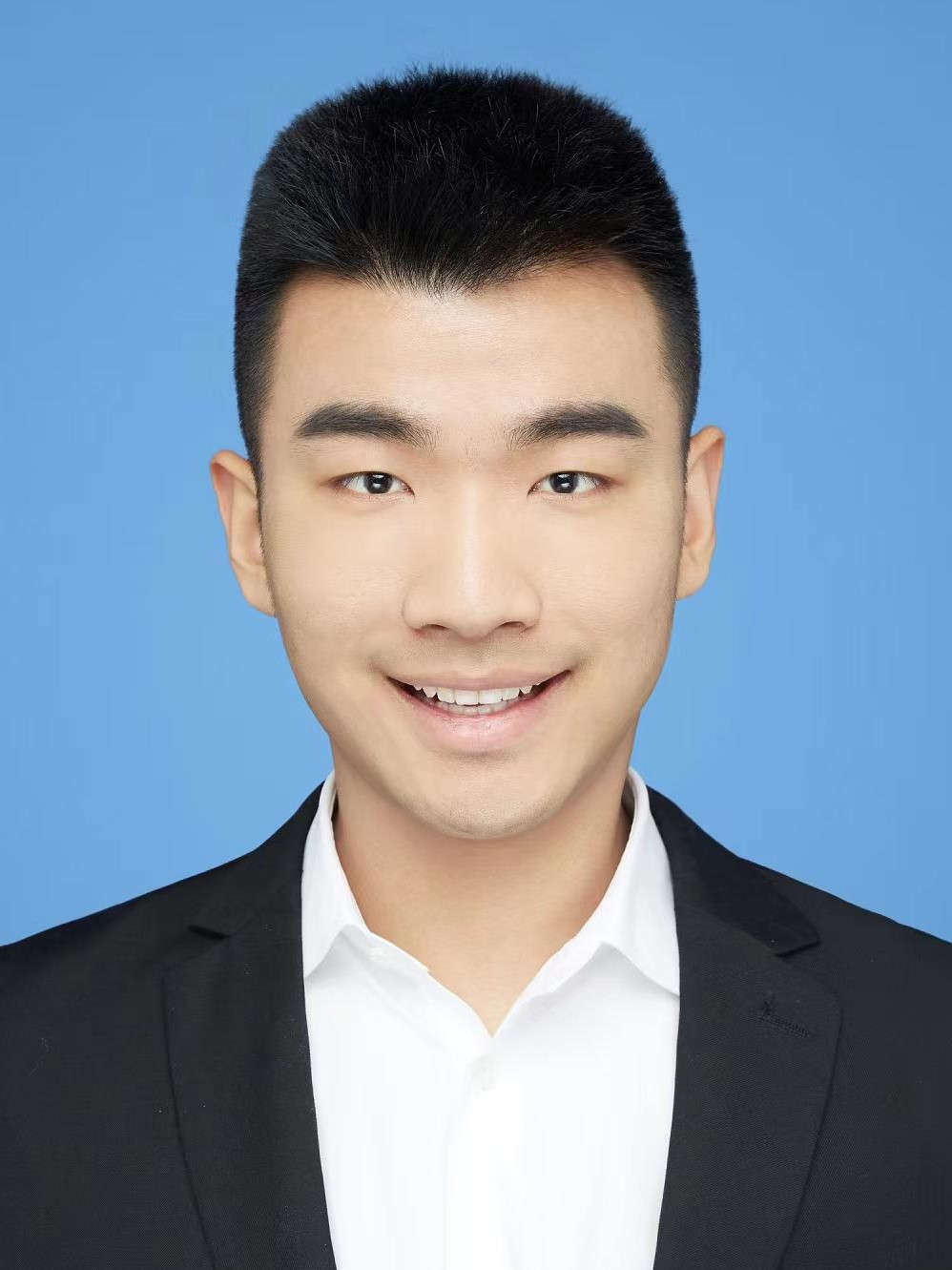}}]{Bowei Xu}
received the B.E. degree from the School of Electronic Engineering, Beijing University of Posts and Telecommunications in 2019 and graduated from the School of Information and Communication Engineering, Beijing University of Posts and Telecommunications with a M.S degree in 2022. He is currently pursuing the Ph.D. degree with the School of Electronic Science and Engineering.
\end{IEEEbiography}


\begin{IEEEbiography}[{\includegraphics[width=1in,height=1.5in,clip,keepaspectratio]{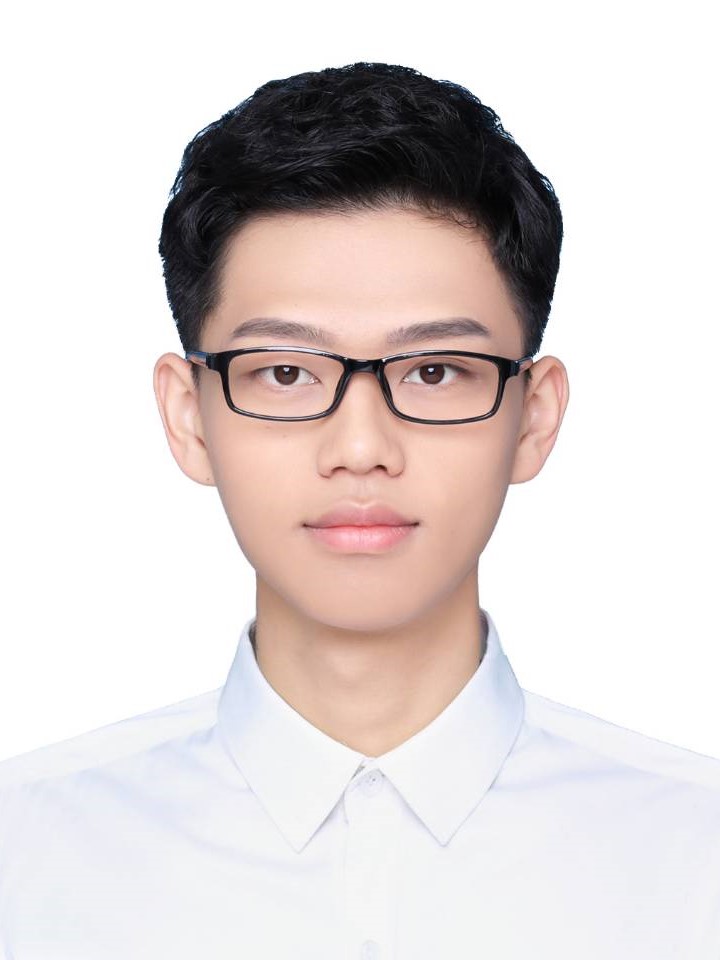}}]{Zicheng Zhang}
Zicheng Zhang received the B.E. degree from the School of Electronic and Information Engineering, Soochow University, in 2021. He is currently pursuing the M.S. degree with the School of Electronic Science and Engineering, Nanjing University.
\end{IEEEbiography}


\begin{IEEEbiography}[{\includegraphics[width=1in,height=1.5in,clip,keepaspectratio]{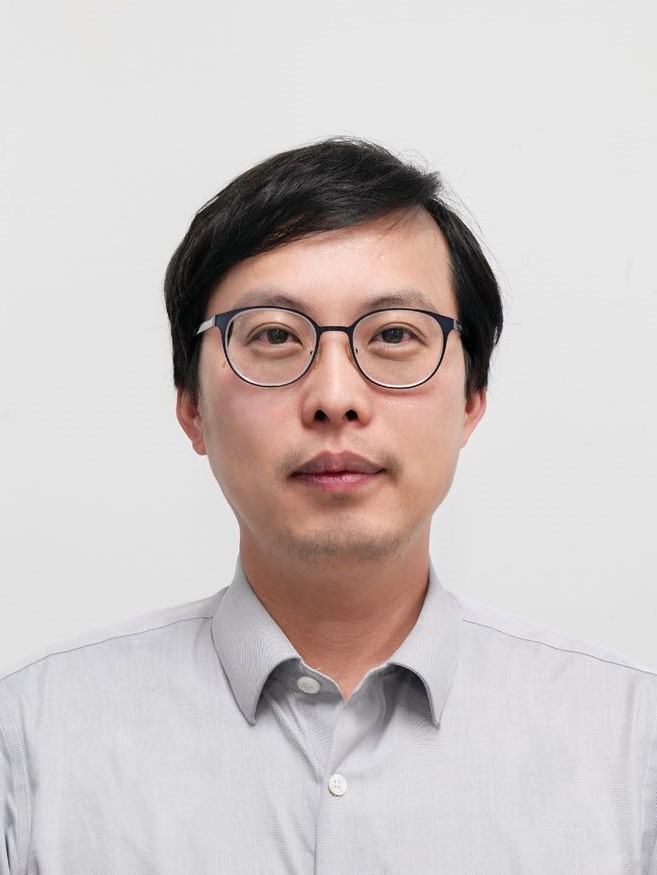}}]{Zhan Ma}
(Senior Member, IEEE) received the B.S.and M.S. degrees from the Huazhong University of Science and Technology, Wuhan, China, in 2004 and 2006, respectively, and the Ph.D. degree from the New York University, New York, NY, USA, in 2011. He is currently a Full Professor with the School of Electronic Science and Engineering, Nanjing University, Nanjing, China. From 2011 to 2014, he has been with Samsung Research America, Dallas, TX, USA, and Futurewei Technologies, Inc., Santa Clara, CA, USA, respectively. His research interests include the learnt image/video coding, and computational imaging. He was the recipient of the 2018 PCM Best Paper Finalist, 2019 IEEE Broadcast Technology Society Best Paper Award, and 2020 IEEE MMSP Grand Challenge Best Image Coding Solution.
\end{IEEEbiography}

\vfill

\end{document}